\documentclass[a4paper,11pt]{article}
\pdfoutput=1 
\usepackage{jcappub} 
\usepackage{amsmath, mathrsfs, amssymb,amsfonts,amsthm,graphicx, epsf, dcolumn, yfonts}
\usepackage{booktabs}
\usepackage{array} 
\usepackage{graphicx}
\usepackage{subcaption}
\usepackage{changepage}
\usepackage{orcidlink}

\definecolor{MyDarkRed}{rgb}{0.71,0.14,0.07}
\definecolor{MyDarkBlue}{rgb}{0.01,0,0.7}

\def\Z#1{_{\lower2pt\hbox{$\scriptstyle#1$}}}

\def\lsim{\mathop{\hbox{${\lower3.8pt\hbox{$<$}}\atop{\raise0.2pt\hbox{$\sim$}}$}}}

\def\dd{\mathop{\text{d}\!}}

\newcommand{\LCDM}[0]{$\Lambda$CDM\,}

\newcommand{\rev}[1]{\color{black}{#1}}

\begin{document}
\bibliographystyle{JHEP}

\title{An effective $\boldsymbol{\Lambda}$-Szekeres modelling of the local Universe with Cosmicflows-4}

\author[a,1]{Marco Galoppo\orcidlink{0000-0003-2783-3603}\note{Corresponding author.},}
\author[b]{Leonardo Giani\orcidlink{0000-0001-6778-1030},}
\author[a]{Morag Hills\orcidlink{},}
\author[c,d]{Aur\'elien Valade\orcidlink{0009-0002-5203-5128}}

\affiliation[a]{School of Physical \& Chemical Sciences, University of Canterbury, \\ Private Bag 4800, Christchurch 8140, New Zealand}
\affiliation[b]{School of Mathematics and Physics, The University of Queensland, \\ Brisbane, QLD 4072, Australia}
\affiliation[c]{Aix Marseille Universite, CNRS/IN2P3, CPPM, \\Marseille, France}
\affiliation[d]{Leibniz Institut fur Astrophysik Potsdam (AIP), \\ An der Sternwarte 16, D-144 Potsdam, Germany}

\emailAdd{marco.galoppo@canterbury.ac.nz}
\emailAdd{l.giani@uq.edu.au}
\emailAdd{morag.hills@pg.canterbury.ac.nz}
\emailAdd{avalade@aip.de}

\abstract{
We develop an effective description of the local cosmic environment --- namely, for redshift $z \lesssim 0.1$ --- to quantify the bias induced by local structure on cosmological observables. Our approach models the metric of the nearby Universe as a superposition of multi-structured $\Lambda$-Szekeres patches, calibrated against the HAMLET peculiar velocity and density field reconstructions of Cosmicflows-4. From this framework we compute the fully inhomogeneous and anisotropic quasilocal expansion field predicted by our model, and use it to assess the impact of local structure on estimates of $H_0$. For this purpose we analyse low-redshift Type Ia supernovae from the Pantheon+ catalogue. We find that accounting for the local structure increases the Hubble tension, yielding a shift in the best-fit value of the Hubble constant of order $\Delta H_0 \approx 0.5\ \mathrm{km\,s^{-1}Mpc^{-1}}$.
}

\maketitle
\section{Introduction}\label{sec:Intro}
The $\Lambda$ Cold Dark Matter ($\Lambda$CDM) model has served as the standard cosmological model for a quarter of a century, providing a fundamental framework for interpreting observations in modern cosmology. One of the pillars of the \LCDM model is the \emph{Cosmological Principle} (CP), i.e., the assumption that on sufficiently large scales the Universe is statistically homogeneous and isotropic. Within the framework of General Relativity (GR), these assumptions are encoded in the Friedmann-Lemaître-Robertson-Walker (FLRW) metrics, which are thus assumed to provide a good description of the average large-scale geometry of the Universe. 

However, as we enter the era of precision cosmology, foundational assumptions of the \LCDM model --- e.g., the validity of the CP, and the interpretation of dark energy as a cosmological constant --- can now be rigorously tested against observational data. Indeed, astrophysical data have become sufficiently precise to study across-the-sky fluctuations and redshift dependence in various cosmological observables.

In this regard, the recent combination of data on supernovae type Ia (SNIa) and baryonic acoustic oscillations (BAO) from DES and DESI with cosmic microwave background (CMB) observations shows a strong preference for dynamical dark energy over a cosmological constant interpretation~\cite{Abbott_2024,Abdul_Karim_2025,Adame_2025}. Furthermore, evidence for anisotropic Hubble diagrams~\cite{Rameez_2021,Cowell_2023,Mc_Conville_2023,Perivolaropoulos_2023,Sorrenti_2023,Tang_2023,Bengaly_2024,Hu_2024,Rameez_2025} point to a significant discrepancy between the dipole anisotropy of the CMB and other distant sources~\cite{Secrest_2022,Dam_2023,Wagenveld_2023}, further challenging the validity of the CP (see~\cite{Aluri_2023} for a review). Finally, high-precision studies of matter bulk flows in the local Universe~\cite{Whitford_2023,Watkins_2023,Watkins_2025} might also indicate tensions with respect to\ the \LCDM predictions.

These studies naturally suggest that the \LCDM model may no longer provide sufficient accuracy when confronted with modern observational data. It is then important to emphasise that any cosmological inference we conduct is from within a highly inhomogeneous and anisotropic environment, embedded in the large-scale structures (LSS) of our Universe. Fittingly, over the past two decades, there has been growing interest within the cosmological community in developing methods to quantify the impact of an inhomogeneous, anisotropic environment on both the observer and the background geometry~\cite{Ellis_1984,Ellis_1987,Buchert_2007,Wiltshire_2011,Clarkson_2011b,Buchert_2012,Koksbang_2019,Koksbang_2020a,Koksbang_2020b,Schander_2021}. 

The effect of the presence of LSS on the large-scale background geometry and dynamics is commonly referred to as \emph{backreaction}. It is conventionally quantified through the Buchert averaging scheme~\cite{Buchert_2000,Buchert_2001,Buchert_2006,Buchert_2018,Buchert_2020}, under the working hypothesis that the FLRW metric well describes the averaged geometry on sufficiently large scales. In this context, backreaction appears as an effective modification of the Friedmann equations, introduced via additional source terms associated with the presence of nonlinear structures within the Universe~\cite{Buchert_2000, Buchert_2001,Buchert_2006,Buchert_2018,Buchert_2020}. Alternatively, mosaic universes --- i.e., Swiss-Cheese models~\cite{Einstein_1945,Marra_2007,Biswas_2008}, pancake solutions~\cite{Najera_2020}, slab patchworks~\cite{Hellaby_2012}, and cellular cosmologies~\cite{Lindquist_1957, Clifton_2012} --- give a direct realisation of statistically homogenous universes which still display some LSS. Within these models, an emergent large-scale background geometry, in agreement with the Buchert averaging scheme predictions, is then recovered and can be directly compared to cosmological observations.

At this present moment, no consensus has emerged within the scientific community on the effective impact of backreaction on cosmological observables~\cite{Clarkson_2011a, Kenworthy_2019, Clifton_2019}. Nonetheless, backreaction effects have been shown to correctly account for the dynamical behaviour of dark energy~\cite{Giani_2025a,Giani_2025b}, and can even entirely replace the role of the latter in explaining the inferred acceleration of the Universe~\cite{Wiltshire_2007,Seifert_2024}. However, although a promising explanation for dynamical dark energy, backreaction effects from LSS cannot account for the anisotropies and small-scale inhomogeneities present in the local Universe.

To model the impact of the local cosmic environment, we would ideally employ a model-independent framework such as the generalised cosmographic expansions developed in~\cite{Heinesen_2021,Kalbouneh_2024,Maartens_2024}. However, this approach is complicated by the relatively small convergence radii of the cosmographic expansions, which are associated with the spatial gradients of small-scale inhomogeneities. This overall restricts the reliability of cosmographic distance estimates to low redshifts. Indeed, this limitation has been shown through investigations with general relativistic numerical simulations~\cite{Macpherson_2021,Macpherson_2023}, Newtonian N-body simulation over an FLRW background~\cite{Adamek_2024}, tests within the spherically symmetric Lemaître--Tolman--Bondi (LTB) spacetimes~\cite{Modan_2024}, Newtonian reconstructions of our local environment~\cite{Koksbang_2024}, as well as in $\Lambda$-Szekeres models~\cite{Hills_2026}. Although such an approach can still be employed to reliably study the presence of multipoles within the expansion field of the local Universe~\cite{Kalbouneh_2025}, alternative approaches may ultimately be better suited to assessing the influence of structures in the local Universe on our cosmological inferences.

Another viable option is to directly model the local structures via exact solutions of the Einstein field equations. Such solutions must either naturally asymptote to an FLRW geometry on cosmological scales, or be intended to describe only the local environment and must be matched --- possibly via a transition layer --- to an FLRW model at larger scales~\cite{Krasinski_1997}. Over the past decade, this approach has attracted growing interest within the cosmological community, with a significant focus on spherically symmetric LTB solutions to model the impact of single structures, typically either a void or an overdensity, on cosmological observables~\cite{Camarena_2021,Camarena_2022,Marra_2022,Camarena_2025}. Moreover, anisotropic Szekeres solutions have been employed to describe small-scale single overdensity-void pairs~\cite{Bolejko_2008,Bolejko_2016}, and a triaxial Bianchi I solution, inspired by a general anisotropic gravitational collapse~\cite{Giani_2021} has been implemented for an anisotropic modelling of the super-cluster Laniakea~\cite{Giani_2024a}. 

In this paper, we adopt an analogous strategy, with the aim of obtaining an effective general-relativistic model of the local Universe as inferred from field-level reconstructions based on peculiar velocity (PV) data as tracers of the underlying matter distribution \cite{Valade_2022,Courtois_2023,Valade_2024, Courtois_2025}. In particular, we consider the HAMLET reconstruction~\cite{Valade_2022,Valade_2024} based on the largest compilation of peculiar velocities currently available, namely the Cosmicflows-4 (CF4) catalogue~\cite{Tully_2023}. The local Universe presents high degrees of inhomogeneity and anisotropy, encoded in the presence of multiple structures. Here, we propose that such an intricate network of structures can be described, at first order, by a combination of multistructured $\Lambda$-Szekeres models~\cite{Szekeres_1975a,Szekeres_1975b,Krasinski_1997}. Furthermore, we assume that the large-scale geometry of the Universe is well approximated by an FLRW metric, and thus constrain our effective model by imposing FLRW boundary conditions. We fit multistructured Szekeres models to the peculiar PV and density HAMLET reconstruction of CF4, and extract the relevant dynamical variables. Using this effective description, we map the spatial curvature and expansion field within the local Universe. Additionally, we compute the impact of the inhomogeneous and anisotropic expansion rate within the local Universe on the redshift-distance relation for an observer located in the Milky Way. Finally, we use these corrections to investigate the impact of the local anisotropic expansion on cosmological observables. 

The remainder of this paper is structured as follows: in Sec.~\ref{sec:Fitting} we employ the CF4 velocity and density field reconstructions to fit multi-structured $\Lambda$-Szekeres models to the local Universe. In Sec.~\ref{sec:Inference} we derive the corrections induced by the effective $\Lambda$-Szekeres model for the local structure on comoving distances, and their impact on cosmological observables, with a particular focus on the Hubble parameter, $H_0$. Sec.~\ref{sec:Conc}
is dedicated to the discussion of our findings and conclusions. Finally, in App.~\ref{app:Szekeres} we give a brief review of the formalism employed to describe the $\Lambda$-Szekeres solutions, in App.~\ref{app:structures} we show that the coarse-graining procedure which we will employ in this work maintains the large-scale structures within the CF4 input fields, and in App.~\ref{app:Code} we present the code routine validation employed in this study. 

\section{An effective model for the local Universe}\label{sec:Fitting}

Our aim is to construct an effective description of the local Universe that qualitatively captures the impact of both anisotropies and inhomogeneities. To this end, we propose a toy model in which the cosmic environment is represented as a patchwork of inhomogeneous spherical wedges --- akin to “pizza slices”. Within each wedge, local structures are modelled by a superposition of monopole and dipole components, allowing us to capture both the dominant isotropic contributions and leading-order directional deviations. With such a description, the local Universe behaves analogously to a class I $\Lambda$-Szekeres spacetime filled with dust and a cosmological constant~\cite{Szekeres_1975a,Szekeres_1975b,Sussman_2012, Sussman_2015,Sussman_2016,Sussman_2017, Buckley_2020} (see also~\cite{Celerier_2024} for a recent review of their use in cosmology). Here, we stress that the effective $\Lambda$-Szekeres description employed in this work is not meant to replace the $\Lambda$CDM framework adopted in the CF4 reconstruction with an unrelated background model. Rather, it provides an exact relativistic reinterpretation of the same reconstructed local environment. Indeed, the $\Lambda$-Szekeres solutions are exact solutions of Einstein's equations whose homogeneous limit is precisely the associated $\Lambda$CDM background spacetime (see e.g.,~\cite{S.W.Goode_J.Wainwright_1982, N.Meures_M.Bruni_2011, Peel_2012b} and~\cite{Krasinski_1997} for an in-depth discussion) . In our application, the effective $\Lambda$-Szekeres patches are therefore anchored to the same background expansion history used in the reconstruction and are required to asymptote to it at the domain boundary. Therefore, the inferred nontrivial expansion field does not reflect a change in the global background, but the inhomogeneous and anisotropic general-relativistic kinematics sourced by the reconstructed matter distribution.

In this framework, the dynamics of the local network of overdensities and voids --- described within its rest frame --- is driven by local differential expansion and collapse. The relative motion between any two points arises entirely from the inhomogeneous and anisotropic expansion field, i.e., a priori, there are no additional PVs sourced by local density gradients. However, we stress that describing the reconstructed flow as peculiar velocities on top of an FLRW background, or equivalently as a local inhomogeneous expansion field, are two different but physically consistent descriptions of the same gravitationally-induced relative motions. Within the $\Lambda$-Szekeres interpretation, these motions are not removed; rather, they are encoded within the inhomogeneous expansion field of an exact inhomogeneous spacetime. This then makes the framework particularly convenient for computing direction-dependent redshift--distance corrections within an internally consistent relativistic model, thus enabling a direct estimate of the impact on cosmological inference of the local cosmic neighbourhood.

Finally, we emphasise that within our model we neglect the time evolution of both anisotropies and inhomogeneities, and instead model the local Universe by assuming that the background Hubble parameter is modified by a line-of-sight dependent function. This approximation is justified by the extremely low effective redshift of the map derived from the CF4 data set, where corrections due to time evolution can safely be neglected~\cite{Giani_2024a, Koksbang_2024}. Here, we consider radially-dependent line-of-sight corrections to the background Hubble parameter, which are necessary to capture the local Universe's inhomogeneities. This is in contrast to the homogenous Laniakea model developed in Ref.~\cite{Giani_2024a}, where local inhomogeneities were neglected.

\subsection{Fields reconstruction from the Cosmicflows-4 data}\label{subsec:CF4}

The large-scale structure (LSS) formed by galaxies on scales of dozen to hundreds of Mpc has been first observed in the 1980s by the first systematic redshift surveys~\cite{deLapperent_etal_1986}. Since then, many more redshift surveys have extended in depth, spatial density and sky-covering our knowledge of the LSS~\cite{SDSS_Collaboration_2000,Jones_6dF_2009,DESICollaboration_2025}. Yet, while providing an intuitive picture of the so-called cosmic-web, this approach only maps the distribution of \emph{galaxies} in the Universe, which are known to be biased tracers of the underlying distribution of matter~\cite{Desjacques_2018}. Indeed, galaxies are found only at the peaks of the distribution of baryonic matter and thus do not directly trace dark matter, which is actually constituting the majority of the matter budget on these scales. Forward modelling methods to reconstruct the density and velocity fields from pure redshift data, introduced in~\cite{Kitaura_2009} and currently lead by the consortium around the BORG code~\cite{Jasche_2019, McAlpine_2025} provide very convincing maps, but they necessarily rely on the assumption of a relationship between galaxy-number density and underlying \emph{total}, gravitationally active matter density. 

Considering the galaxies' velocities instead of their positions presents a powerful advantage. Indeed, taken as tracers of the velocity field, galaxies are sensitive to the total distribution of gravitationally active matter, be it luminous or dark. Galaxies' velocities have no bias at leading order. However, observations of peculiar velocities are relatively more challenging, leading to scarce and noisy data prone to interpretation biases~\citep{Strauss_1995,Nusser_2025}. The Wiener-Filter(WR) / Constrained Realizations (CRs) framework~\citep{Hoffman_Ribak_1992,Zaroubi_1999} has been the standard approach for the last few decades, yet two inherent assumptions clash with the reality of the data, namely that (i) observational uncertainties are Gaussian, and (ii) that the positions of the constraints are known. This motivated the application of field-level forward modelling techniques to velocity data. Forward modelling consists of a Monte Carlo exploration of the posterior distribution of an arbitrary complex Bayesian model~\citep{Kitaura_2009,Lavaux_2016,Graziani_2019,Valade_2022, Boruah_2022}. In practice, these models describe the fields in the same manner as the WF/CRs framework, but differ in their modelling of the observations. Forward modelling methods, such as the one employed here, \rev{thus circumvent the use of specialised velocity estimators such as~\cite{Sorce_2015,Hoffman_2021,Watkins_2023,Sorce_2023} that are commonly used to ``correct'' the velocity data prior to the application of the WF.}

Linear reconstruction methods rely on two major assumptions. First, that the fields can be described as Gaussian and that the power spectrum of the density field is that of the CMB~\citep{Planck_2018}. Second, that the matter density perturbation field $\delta$ and the three-dimensional PV field $\vec{v}_\mathrm{pec}$ are related by the linearised continuity equation~\citep{KodamaSasaki_1984,Mukhanov_1992,Malik_2009}
\begin{equation}
\partial_t \delta + \frac{1}{a} \vec{\nabla} \cdot \vec{v}_\mathrm{pec} = 0 \, , \label{eq:cont1}
\end{equation}
where $a(t)$ is the background scale factor. The density contrast field can be factorised as $\delta(\vec{x},t) = D_+(t)\delta_0(\vec{x})$, with $D_+(t)$ representing the linear growth factor of density fluctuations~\citep{KodamaSasaki_1984,Mukhanov_1992,Malik_2009}. Substituting this form into Eq.~\eqref{eq:cont1} leads to a convenient expression connecting the divergence of the PV to the density perturbation
\begin{equation}
\vec{\nabla} \cdot \vec{v}_\mathrm{pec} + a H f \delta = 0 \, , \label{eq:cont2}
\end{equation}

This assumption, commonly referred to as \emph{linear theory}, is strictly valid only on large scales, of order tens of megaparsecs, although it is often extrapolated down to scales of a few megaparsecs. Applied outside its range of validity, linear theory may yield locally negative density in void regions and tends to misestimate the contrast of structures in dense regions. 

This work is based on density and velocity fields obtained from a linear field-level forward-modelling approach, namely the HAMLET reconstruction framework~\cite{Valade_2022,Valade_2024}. 
\rev{The physical model of HAMLET is similar to that presented by~\cite{Graziani_2019}, but the Monte Carlo Gibbs sampling is replaced by a Hamiltonian Monte Carlo sampler~\cite{Neal_2011}. This sampler is better suited to highly-dimensional problems and has GPU-accelerated exploration, resulting in a speed-up of approximately $10{,}000$. The present reconstruction is an application of HAMLET} to the largest compilation of peculiar-velocity measurements available to date, namely the CF4 catalogue~\cite{Tully_2023}. The dataset comprises more than $56{,}000$ peculiar velocities, providing nearly all-sky coverage out to redshift $z \sim 0.05$ (namely, a background comoving distance of $\sim 150~\mathrm{Mpc}/h$), with particularly good sampling in the Galactic and celestial northern hemispheres extending to measured redshifts of $z = 0.1$, i.e., $\sim 300~\mathrm{Mpc}/h$, with some sparse data points with larger measured luminosity distances of up to $\sim 500~\mathrm{Mpc}/h$~\cite{Tully_2023}. The data is grouped in $38{,}000$ groups which mitigates the non-linear effects in the data and reduce the uncertainty on the constraints. \rev{Additionally, a field-level, uncorrelated noise of amplitude $\sigma_{\rm NL} = 300\,{\rm km/s}$ is introduced to further reduce the remaining tension between the observations and the linear framework of the reconstruction scheme. The introduction of this parameter is standard and also present in WF-based reconstructions. By construction, the adopted Bayesian framework transitions smoothly from a data-driven reconstruction in regions well sampled by observations to a prior-driven regime elsewhere in the box. In the absence of constraining data, the mean density contrast and PV fields converge to zero, while the scatter around the mean spans the full range allowed by \LCDM{}. On the one hand, this makes HAMLET's predictions conservative, as the reconstruction is by default dictated by the prior model, here \LCDM{}, while any deviation from the prior-predicted behaviour directly stems from the data. On the other hand, the model correctly reflects possible tension with \LCDM{} within the data. For instance, if the data yields a bulk flow that is large (or low) for \LCDM{}, the reconstruction would reflect it in the central region but converge anyway to \LCDM{} where the data is weak. Forward models that measure the bulk flow without a \LCDM{} prior exist, yet they cannot reconstruct field-level information, see e.g.~\cite{Duangchan_2025}.}

Finally, the HAMLET reconstruction evaluates the density and velocity fields of a regular grid of $L_{\rm box} = 1\,{\rm Gpc/h}$ side length divided into 256 nodes, leading to a spatial resolution of $4\,{\rm Mpc/h}$. Hence, the reconstructed volume is a priori sufficiently large enough to encompass the scales relevant to large-scale bulk flows~\cite{Whitford_2023,Watkins_2023,Watkins_2025}, the major cosmic structures (i.e, walls, filaments and voids) within our cosmic neighbourhood (see e.g.,~\cite{Tully_2014,Pomarede_2015,T.Shanks_etal_2019,Wong_2022}), as well as various measures of homegeneity scale (for example~\cite{Hogg_2005,Pandey_2013,Pandey_Sarkar_2015,Dias_2023,Sylos_Labini_2026,L.Giani_etal_2026}). However, we note that the assumed periodicity of the box prevents the fitting of modes larger than the box-size, hence, some large scale power is lost. This effect is known to artificially dampen of $20\%$ (respectively $50\%$) the amplitude of the large scale bulk flow at $L_{\rm box} / 10 = 100\,{\rm Mpc}/h$ (respectively $L_{\rm box}$)~\cite{Duangchan_Prep}. Thus, we caution that the inferred results should only be compared to model predictions in the same volume. Finally, we note that within the HAMLET reconstruction $\Omega_M = 0.3$, and $H_0 = 74.6~\mathrm{km\,s}^{-1}\,\mathrm{Mpc}^{-1}$~\cite{Valade_2024}, i.e., the best fit cosmological parameters extracted for the CF4 data set~\cite{Tully_2023}. Thus, no inconsistency bias is introduced at the level of background model parameter between data and reconstruction.

\subsection{Coarse-grained spatial geometry}\label{subsec:CGSG}

The starting point for our analysis is to derive an effective description of the spatial geometry of the nearby cosmic environment. To this end, we partition the CF4 map into a set of spherical wedges, each further subdivided into radial shells. Specifically, we divide the sky into 256 angular regions by adopting 16 bins in both the azimuthal and polar directions. In addition, we introduce 20 radial bins, defined such that the volume between successive shells increases uniformly, i.e., each radial bin spans an equal comoving volume. Such angular partitioning allows one to properly capture local anisotropies, whilst the radial binning permits an accurate description of the radial dependence of the inhomogeneities. 

In the $\Lambda$-Szekeres spacetimes, the dipole moment within the local matter distribution is encoded within via a single dipole function, $\boldsymbol{W}(r,\theta,\phi)$ (see App.~\ref{app:Szekeres} and, e.g.,~\cite{Szekeres_1975b,Sussman_2012, Sussman_2015,Sussman_2016,Sussman_2017, Buckley_2020}). The dipole function is conventionally expressed as
\begin{equation}
    \boldsymbol{W}(r,\theta,\phi) = -X(r)\cos(\phi)\sin(\theta) - Y(r)\sin(\phi)\sin(\theta) - Z(r)\cos(\theta)\, ,
\end{equation}
where $X(r),\, Y(r),$ and $Z(r)$ are arbitrary functions of the radial coordinate. We follow the ansatz of~\cite{Sussman_2015,Sussman_2016} for their functional form, i.e., in each $i$-th spherical wedge we take these functions to be piecewise defined over each $j$-th spherical shell by
\begin{align}
    &X_{ij}(r) = -\cos(\phi_{ij})\sin(\theta_{ij})f_{j}(r) \, , \\
    &Y_{ij}(r) = -\sin(\phi_{ij})\sin(\theta_i)f_{j}(r)\, ,\\
    &Z_{ij}(r) = -\cos(\theta_{ij})f_{ij}(r)\, ,
\end{align}
where in each spherical shell $(\theta_{ij},\phi_{ij})$ are the angular coordinates of the density maxima within the $j$-th spherical shell in the  
$i$-th spherical wedge. We have also defined the radial functions $f_{j}$ as 
\begin{equation}
    f_{j}(r) = \sin^2\left[\left(\frac{r-r_*^{j-1}}{r_*^{j}-r_*^{j-1}}\right)\pi\right]\, ,
\end{equation}
where $r_*^j$ is the outer boundary of the $j$-th spherical shell. 

We emphasise that describing the local Universe through a spherical wedge-partition of multistructured $\Lambda$-Szekeres models can only, at best, serve as an effective approximation. This construction inherently introduces discontinuities at the interfaces between wedges --- namely, layers of physical matter determined by the density gradient across the matching surfaces (see, e.g.,~\cite{Sussman_2016}). However, we note that: (i) such discontinuities are already an unavoidable feature of any coarse-grained description of the local Universe; and (ii) in our approach, we neglect the time evolution of the nearby cosmic environment, so these artifacts ultimately play no role in our investigation. Indeed, the effective model we construct is intended only to probe how local spacetime expansion affects cosmological inferences. It is not meant to serve as a dynamical model of the local Universe's evolution.

\subsection{Effective \texorpdfstring{$\boldsymbol{\Lambda}$}{Λ}-Szekeres reconstruction}
\label{subsec:ESR}
The matter density $\rho$, the expansion rate scalar $H/3$, and the three-dimensional spatial curvature $\mathcal{K}$ in a $\Lambda$-Szekeres model are expressed within the quasilocal scalar formalism originally developed by Sussman and Bolejko~\cite{Sussman_2012}, and further extended in Refs.~\cite{Sussman_2015,Sussman_2016,Sussman_2017}, as
\begin{align}
    &\rho = \rho_q + D^{(\rho)} \, ,\\
    &H = H_q + D^{(H)} \, ,\\
    &\mathcal{K} = \mathcal{K}_q + D^{(\mathcal{K})} \, ,
\end{align}
where $\rho_q(t,r)$, $H_q(t,r)$, and $\mathcal{K}_q(t,r)$ denote proper volume averages taken over spherical domains centred at the origin of the corresponding physical scalars (see App.~\ref{app:Szekeres} and~\cite{Sussman_2012, Sussman_2015,Sussman_2016,Sussman_2017}). In this sense, the coordinate origin is thus interpreted as the observer-centred anchor of the effective reconstruction. Then, since the local Universe is inhomogeneous and anisotropic, shifting this origin (i.e., the observer in our framework) would in general lead to a different radial decomposition and hence to different quasilocal averages. Therefore, the present construction should be understood as an effective description of the cosmic neighbourhood as reconstructed from our observational position. These quasilocal scalars then define a reference LTB background --- able to capture the impact of even strong inhomogeneities --- with respect to which the quantities $D^{(\rho)}$, $D^{(H)}$, and $D^{(\mathcal{K})}$ represent exact deviations that encode local anisotropies. The quasilocal scalars and their respective exact deviations satisfy the constraints~\cite{Sussman_2012, Sussman_2015,Sussman_2016,Sussman_2017}
\begin{align}
    & H_q^2 = \frac{8\pi G}{3}\rho_q - \mathcal{K}_q +\frac{\Lambda}{3} \, ,\label{eq:cs1}\\
    &  D^{(\mathcal{K})} =  \frac{8\pi G}{3}D^{(\rho)} - 2H_qD^{(H)} \, . \label{eq:cs2}
\end{align}
Here, Eq.~\eqref{eq:cs1} serves as the analogue of a Friedmann equation for an FLRW background independently defined on each spherical shell, whilst Eq.~\eqref{eq:cs2} mirrors the linear constraint relating curvature, density, and expansion perturbations in standard cosmological perturbation theory. However, in contrast with linear perturbation theory, the exact deviations of the quasilocal formalism do not need to be small and can thus encode strong anisotropies.

Here, we aim to leverage the quasilocal formalism of Szekeres models to build an effective description of the impact of the local LSS on cosmological inferences. To this end, we make use of data from the HAMLET CF4 reconstruction, and employ both the density and three-dimensional PV fields. As previously discussed, we assume that the inferred PVs arise solely from the anisotropic and inhomogeneous expansion field in the nearby cosmic environment. The key model-building assumption is thus the existence of a line-of-sight dependent function that encodes the directional variation in the local expansion rate, $H_{\mathrm{los}}(\theta,\phi,z)$, such that
\begin{equation}
    H_{\mathrm{los}}(\theta,\phi,z) \approx c\frac{z}{\bar{d}_L}\, ,
\end{equation}
where $z$ is the measured redshift, corrected for the peculiar motion of the observer, $v_0$, i.e., $1+z = (1+z_{\mathrm{obs}})(1+v_0/c)$, and $\bar{d}_L$ is the luminosity distance associated to the measured redshift in the fiducial cosmology. In other words, we are assuming that the perceived relative motion between objects in the local Universe is given by the spatial variation of the local Hubble expansion, rather than by intrinsic motions or external tidal fields, following
\begin{equation}
    v_\mathrm{pec} \approx H_{\mathrm{los}}\left(\frac{\bar{d}_L}{1+z}\right) \, ,
\end{equation}
where we now identify $H_{\mathrm{los}}=H-H_{|r=0}$, with $H$ being the Hubble field of the $\Lambda$-Szekeres models, and $r = 0$ being the observer position. Moreover, we single out the large-scale background contribution by writing 
\begin{equation}
    H = H_0 +\Delta H\, ,
\end{equation}
where $H_0$ is the underlying Hubble parameter of the fiducial FLRW cosmology employed in the reconstruction. We thus obtain 
\begin{equation}
    v_\mathrm{pec} \approx \left(\Delta H - \Delta H_{|r=0}\right)\left(\frac{\bar{d}_L}{1+z}\right) \, .
\end{equation}

Therefore, from the PV data of CF4, we can directly extract the Hubble field, $H$, of the underlying $\Lambda$-Szekeres models, given an estimate of the expansion rate at the observer's location. Here, we will consider as ansatz $\Delta H_{|r=0} = 0$, i.e., we take the expansion within our cell to be, on average, the same as the fiducial cosmological models. This assumption --- adopted consistently throughout the analysis --- is motivated by the fact that the HAMLET reconstruction provides PVs in the comoving frame. As a result, the density and PV fields used to calibrate the effective $\Lambda$-Szekeres model are not affected by a dipole bias induced by the peculiar motion of the observer with respect to the Hubble flow.

Constructing a description in terms of quasilocal scalars and their exact deviations also requires determining the proper volume of the reconstructed cells, which is necessary for performing the weighted integration defining each quasilocal quantity. To determine the proper volume of the reconstructed cells, we exploit the linear nature of the employed CF4 reconstruction. Indeed, in this case we can find the proper volume, $\dd V_\mathrm{prop}$ as given by
\begin{equation}
    \dd V_\mathrm{prop} \simeq a_0^3\left(1+\delta-\boldsymbol{W}\right)^{1/2} \dd V_\mathrm{flat}\, ,
\end{equation}
where $\dd V_\mathrm{flat}$ is the volume element in flat space, $\delta$ is the linear density contrast, and $a_0(r)$ is the current-time seed LTB scale-factor, which can be determined, given FLRW boundary conditions, to be~\cite{Sussman_2015,Sussman_2017}
\begin{equation}
a_0(r) \simeq \exp\left[-\int_r^{r_\mathrm{FLRW}}\frac{\delta_{\mathrm{mon}}(r')}{r'}\dd r'\right]\, ,
\end{equation}
where $\delta_{\mathrm{mon}}$ is the monopole term in the density contrast, and $r_\mathrm{FLRW}$ represents the radius at which FLRW boundary conditions are imposed. We thus find that the proper volume of each cell can be determined once the dipole function and the density contrast are known. 

\subsection{Quasilocal expansion field of the cosmic neighbourhood}\label{subsec:q}

An intrinsic limitation of existing velocity field reconstructions is due to small scale nonlinear motions and gravitational clustering, which can only be partially modelled in the reconstruction process\footnote{%
For example by adding a constant nonlinear component to the velocity dispersion which is treated as a free parameter~\cite{Courtois_2023,Valade_2024,Courtois_2025}.
} unavoidably contaminating the linear reconstruction. As a consequence, for example, there are multiple shells in which the reconstructed \emph{linear} density contrast is $\delta \geq 6$ (with a probability of occurring for a single cell of order $10^{-4} \%$), or even below the minimum $\delta < -1$. Therefore, to construct a multistructured $\Lambda$-Szekeres effective model of the local Universe that leverages the linear (and quasi-linear\footnote{%
As proven in~\cite{GasparBuchert_2021, GasparBuchertOstrowski_2023}, the $\Lambda$-Szekeres models are exact solutions of the first-order scheme of the Generalised Relativistic Zel'dovich Approximation (GRZA). Therefore, since the GRZA --- similar to its Newtonian counterpart --- remains valid beyond the linear regime of the density contrast at first order, we can expect a first-order $\Lambda$-Szekeres approach to also provide a reliable description within the quasi-linear regime.
}) regime of the HAMLET CF4 reconstruction, we must further process the data to mitigate nonlinear contamination. To achieve this, we employ a smoothing algorithm that computes averages over groups of neighboring cells for both the density contrast and PV fields, thereby filtering out small-scale nonlinear fluctuations while retaining the large-scale structure relevant to our analysis. Specifically, we apply a coarse-graining procedure over groups of $16\times16\times16$ voxels --- equivalent to an effective coarse graining scale of $\simeq 64~\mathrm{Mpc/h}$ --- which reduces the density contrast to the range $-0.3 \leq \delta \leq 0.7$. This level of smoothing ensures that the reconstructed fields remain within the quasi-linear regime whilst preserving the main morphological features of the CF4 structures (see also App.~\ref{app:structures}). We thus perform the multistructured $\Lambda$-Szekeres fitting over the CF4 coarse-grained reconstruction data following the procedure detailed in subsections~\ref{subsec:CGSG} and~\ref{subsec:ESR}.

Here, our main interest lies within the differential expansion field within our local Universe, which ultimately determines the observed PVs in the CF4 catalogue. Interestingly, we find that within the smoothed CF4 reconstruction, the correction to the Hubble field induced by the exact deviations predicted by the $\Lambda$-Szekeres model $D^{(H)}$ with respect to the quasilocal average $H_q$ are negligible if compared to the corrections induced when considering $H_q$ alone. In particular, if we consider $H = H_0 + \delta H_q + D^{(H)} = H_q + D^{(H)}$, we have $D^{(H)}/H_0 \ll \delta H_q /H_0$\footnote{
Here, we have not specified the cosmological parameters of the underlying background. Thus, the ratios of expansion rates represent the natural observable within the model.
}. Therefore, we can \emph{effectively} neglect the $D^{(H)}$ term within the map when describing the expansion field, as it only generates higher order contributions.

\begin{figure}[t]
\centering
\includegraphics[width=0.6\textwidth]{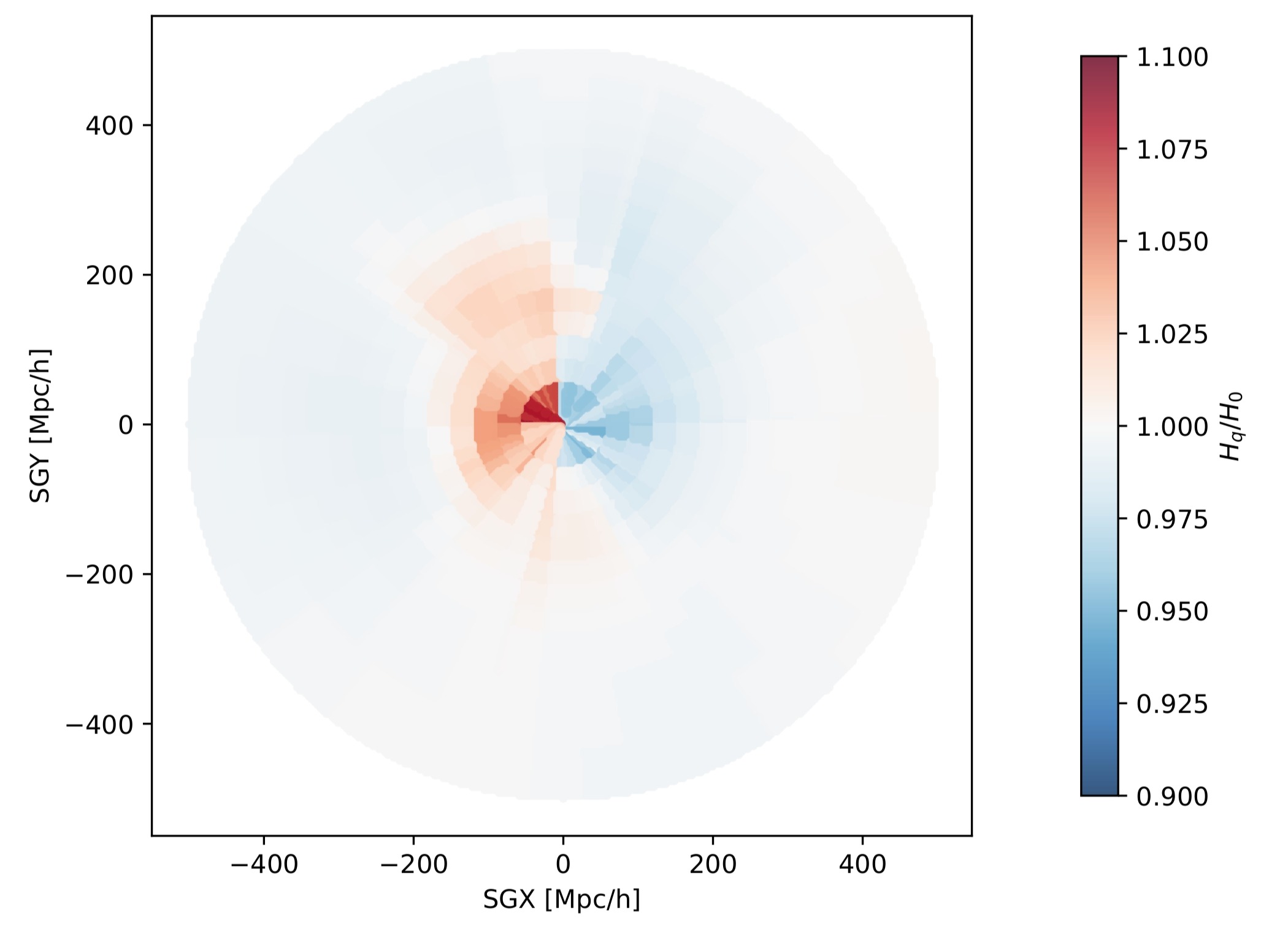}
\includegraphics[width=0.6\textwidth]{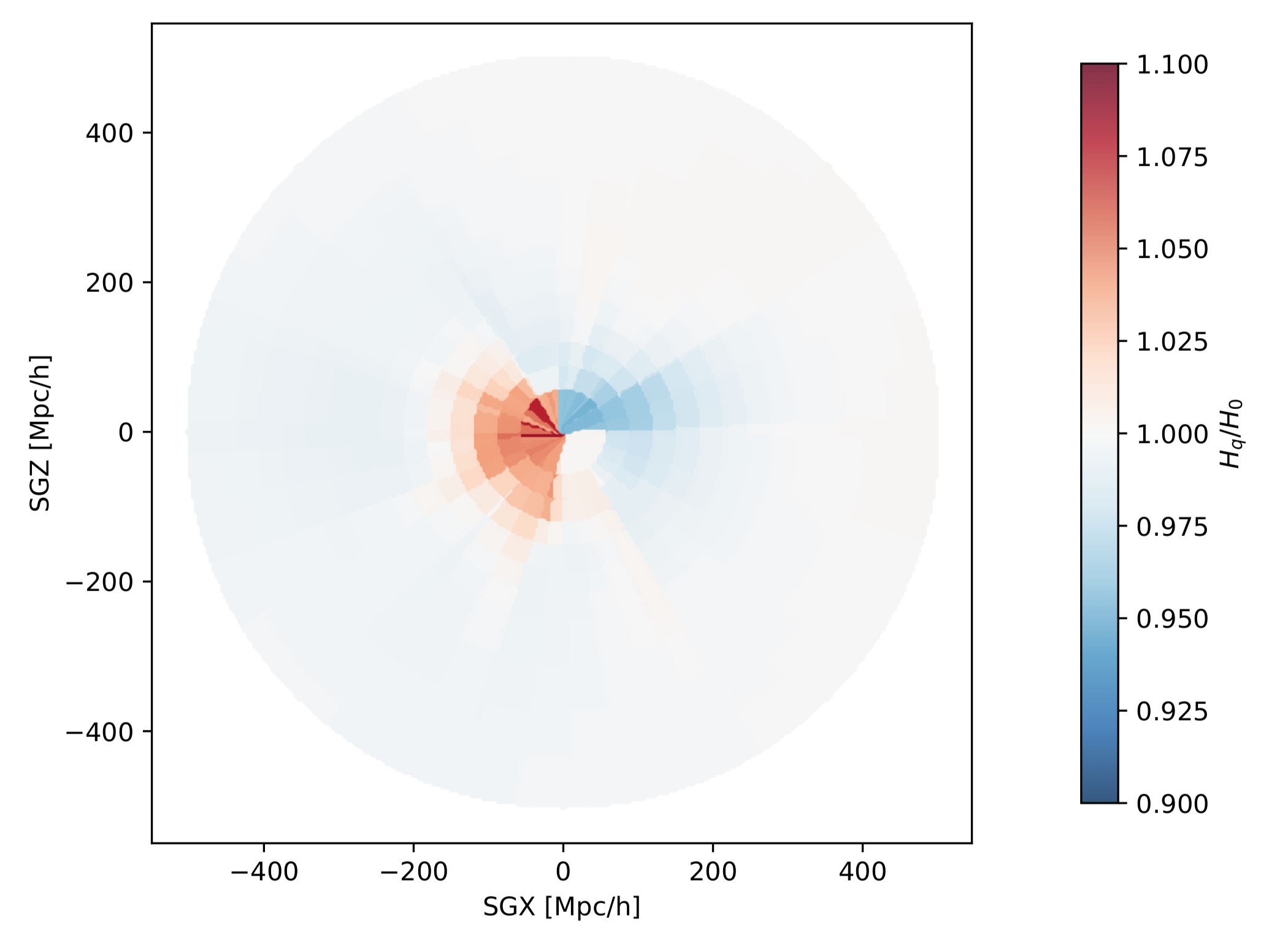}
\includegraphics[width=0.6\textwidth]{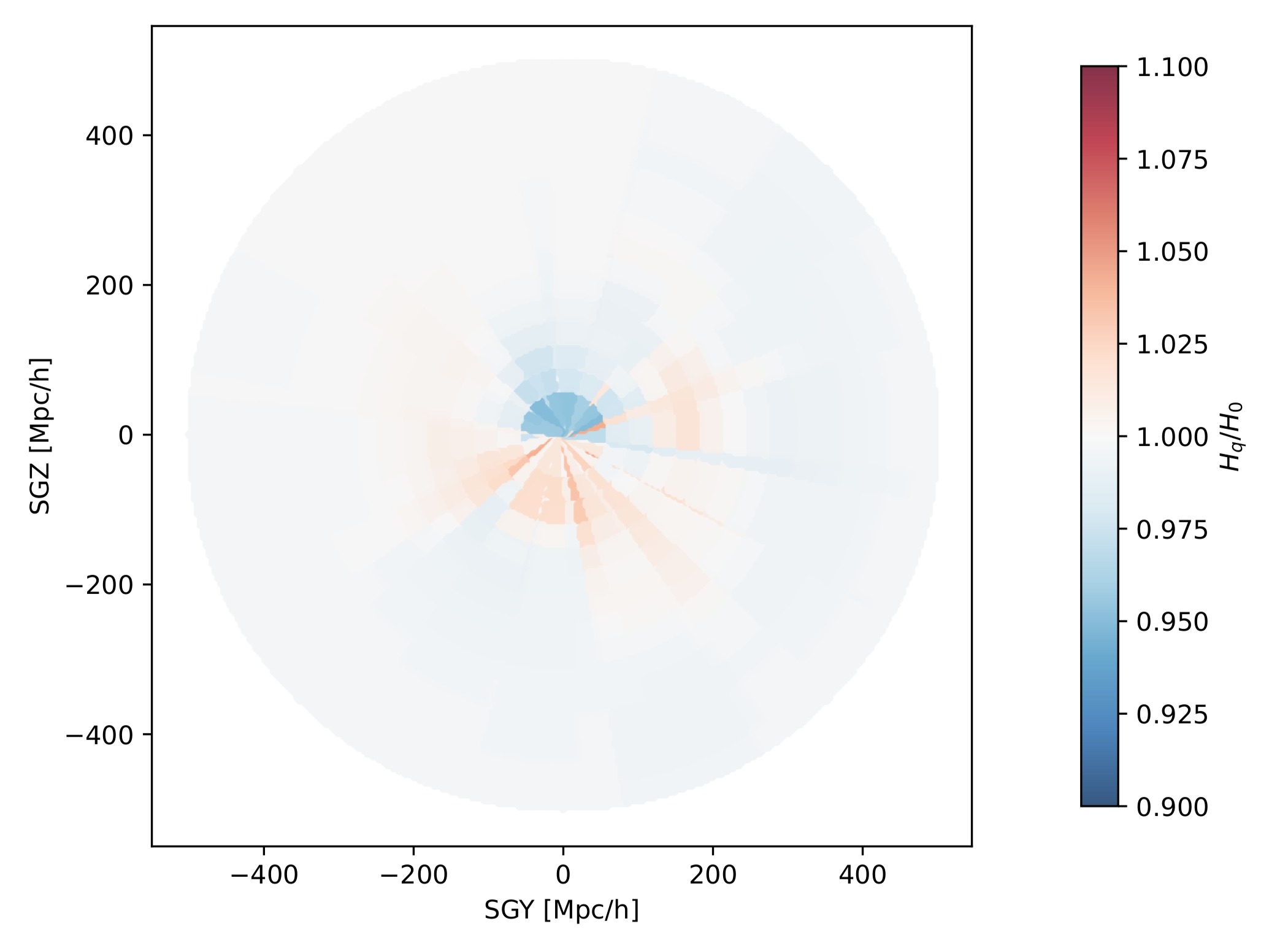}
\caption{Ratio $H_q/H_0$ in the three fundamental planes: SGX--SGY (top), SGX--SGZ (middle), and SGY--SGZ (bottom).}
\label{fig:Hq_all_planes}
\end{figure}

In Fig.~\ref{fig:Hq_all_planes}, we plot the ratio between the reconstructed quasilocal expansion field, $H_q$, and the underlying Hubble parameter, $H_0$, in the three fundamental planes. We find predicted variations in the Hubble field absolute magnitude of up to 10\%, with a non-trivial spatial profile ultimately reflecting the anisotropy and inhomogeneity of the local cosmic web. In particular, along the SGX–SGY and SGX–SGZ planes, we observe an essentially dipole-like feature in the expansion field, which is absent in the SGY–SGZ plane, further highlighting the directional dependence of the local Hubble flow within the local Universe. Moreover, we note that this behaviour is consistent with recent independent analyses reporting a dipole-like pattern in the local expansion field inferred from CF4 data~\citep{Salzano_2025,Kalbouneh_2025}.

We emphasise the quasilocal nature of the reconstructed field, $H_q$, which represents a weighted spatial average of the underlying expansion rate. As a result, the correspondence between $H_q$ and the actual matter distribution is spatially smoothed, leading to a local offset between the position of a structure and the associated variation in $H_q$. For instance, a void would induce higher $H_q/H_0$ values, but the corresponding maximum typically occurs displaced from the void’s geometric centre due to the spatial averaging and binning processes. Therefore, although a correspondence between the features of $H_q$ and the underlying physical structures exists, caution must be exercised when interpreting this relation directly.

\begin{figure}[htbp]
    \centering  
    \includegraphics[width = 0.6\textwidth]{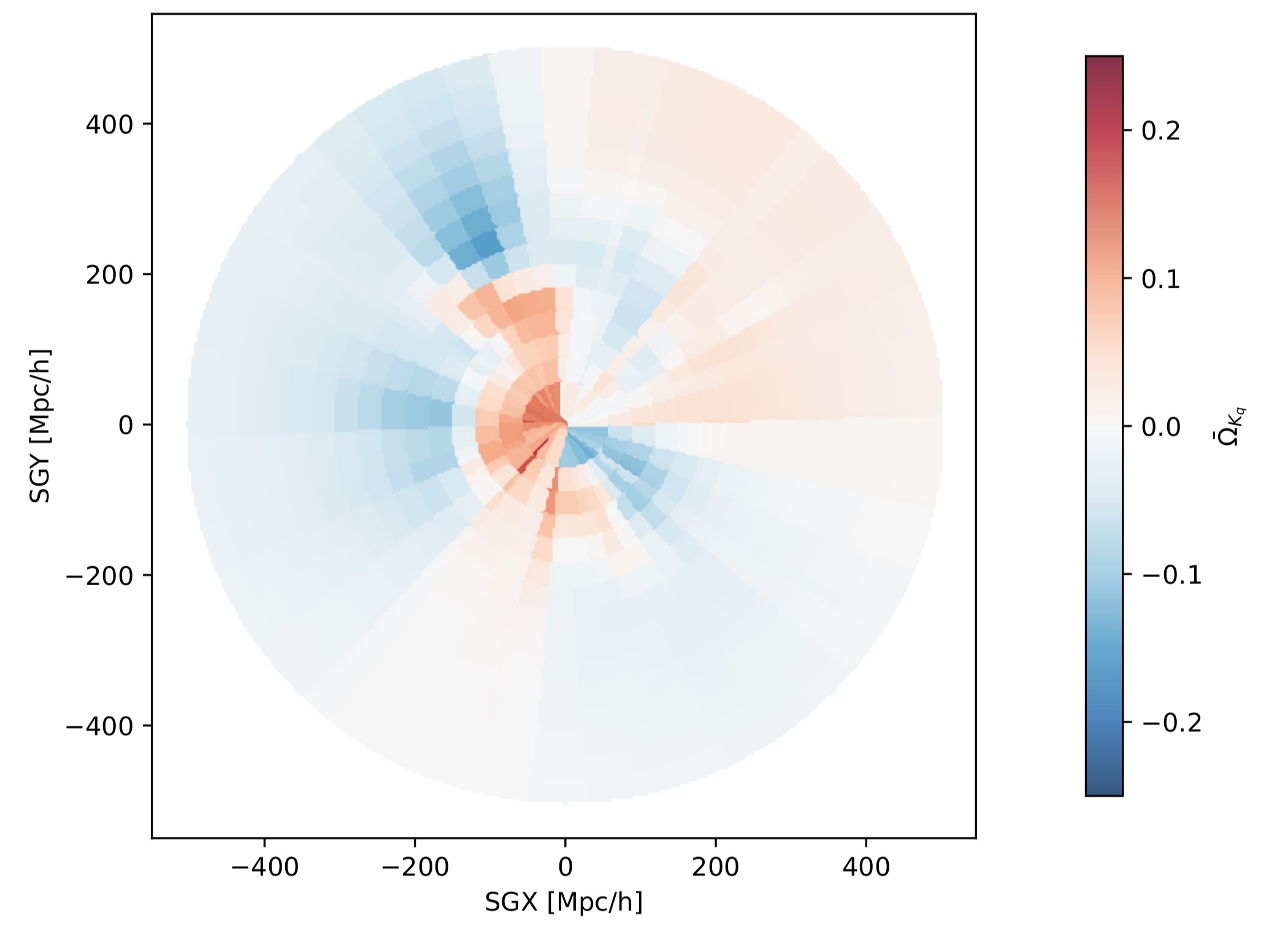}
    \caption{The spatial profile and magnitude of $\bar{\Omega}_{Kq}$ on the SGX-SGY plane}
    \label{fig:Kq_SGXSGY}
\end{figure}

In this regard, it is of interest to consider the quasilocal average of the spatial curvature, which can be constructed directly from the averaged expansion and density fields following Eq.~\eqref{eq:cs1}. In particular, it is convenient to define $\bar{\Omega}_{\mathcal{K}_q} := -\mathcal{K}_q/H_0^2$, which provides a measure of the energy contribution of spatial curvature with respect to the fiducial cosmological background. In Fig.~\ref{fig:Kq_SGXSGY}, we show, as an example, the distribution of $\bar{\Omega}_{\mathcal{K}q}$ on the SGX–SGY plane. We find a rough correlation between regions of negative curvature ($\bar{\Omega}_{\mathcal{K}_q} > 0$), generally associated with voids, and regions where $H_q/H_0 > 0$. Nonetheless, as expected, an exact spatial correspondence cannot be established within this framework due to the use of quasilocal averages with respect to a central observer. Interestingly, we find non-negligible spatial curvature contributions to the energy budget in several bins, highlighting the possible relevance of spatial curvature in the inference of local cosmic observables, such as bulk flows.
\\

\section{Impact on cosmological inference}\label{sec:Inference}

We are now in a position to assess how the multi-$\Lambda$-Szekeres description of the cosmic environment affects cosmological observations. As a testbed, we use the subsample of Pantheon+ type Ia supernovae (SNe~Ia) in the redshift range $0.023 < z < 0.15$, i.e., the same range employed by the SH0ES collaboration~\cite{Riess_2022} to infer the value of the Hubble constant. 

Note that, a priori, the different local expansion rates predicted by the patchwork of multi-structured $\Lambda$-Szekeres models lead to corresponding anisotropic and inhomogeneous modifications in the distance–redshift relation. Here, following the findings of Sec.~\ref{subsec:q}, we can approximate the luminosity distance-redshift relationship predicted by our model within the volume of the used CF4 reconstruction by exclusively using the quasilocal averages of the expansion field, yielding (see also App.~\ref{app:Code})
\begin{equation}\label{eq:dL}
    d^{\Lambda-\mathrm{Sk}}_L(z,\hat{\mathbf{n}}) \simeq (1+z)\int_0^z\frac{c\dd z}{H_{q}(z, \hat{\mathbf{n}})} \, ,
\end{equation}
where $\hat{\mathbf{n}}$ indicates the line of sight unit vector. We can then consider the global, direction-averaged luminosity distance $\bar d_L(z)$ --- identified with that of the underlying fiducial FLRW model --- as a baseline for comparison against the anisotropic corrections predicted by the effective model. For each SNe~Ia we thus write its predicted luminosity distance within the multi-structured $\Lambda$-Szekeres reconstruction as
\begin{equation}
d_L^{\Lambda-\mathrm{Sk}}(z, \hat{\mathbf{n}}) :=
\bar d_L(z)\Delta d_L^{\Lambda-\mathrm{Sk}}(z, \hat{\mathbf{n}})\,,
\end{equation}
where $\Delta d_L^{\mathrm{Sk}}$ encodes the fractional deviation induced by the anisotropic local geometry along the line of sight.

To quantify the significance of these deviations, we adopt a bootstrapping approach and generate one thousand mock realisations of the CF4 smoothed reconstructions. As the CF4 coarse-grained fields provide a mean value and a $1\sigma$ uncertainty for both the density contrast and PV fields in each voxel, we build an ensemble by sampling these quantities independently voxel-by-voxel (i.e., a Monte--Carlo error propagation at the field level). We then model each CF4 realisation using the proposed multi-$\Lambda$-Szekeres description, and extract a corresponding $H_q$ map, from which we compute the luminosity-distance corrections for the SNe~Ia sample. For each supernova, we then assign the correction to the luminosity distance and its associated uncertainty as the mean, $\bar{\Delta} d_L^{\Lambda\text{-Sk}}$, and standard deviation, $\sigma_{\Delta d_L}^{\mathrm{Sk}}$, respectively, of the corrected distances across all realisations of the HAMLET CF4 reconstructions.

\begin{figure}
    \centering
\includegraphics[width=1\linewidth]{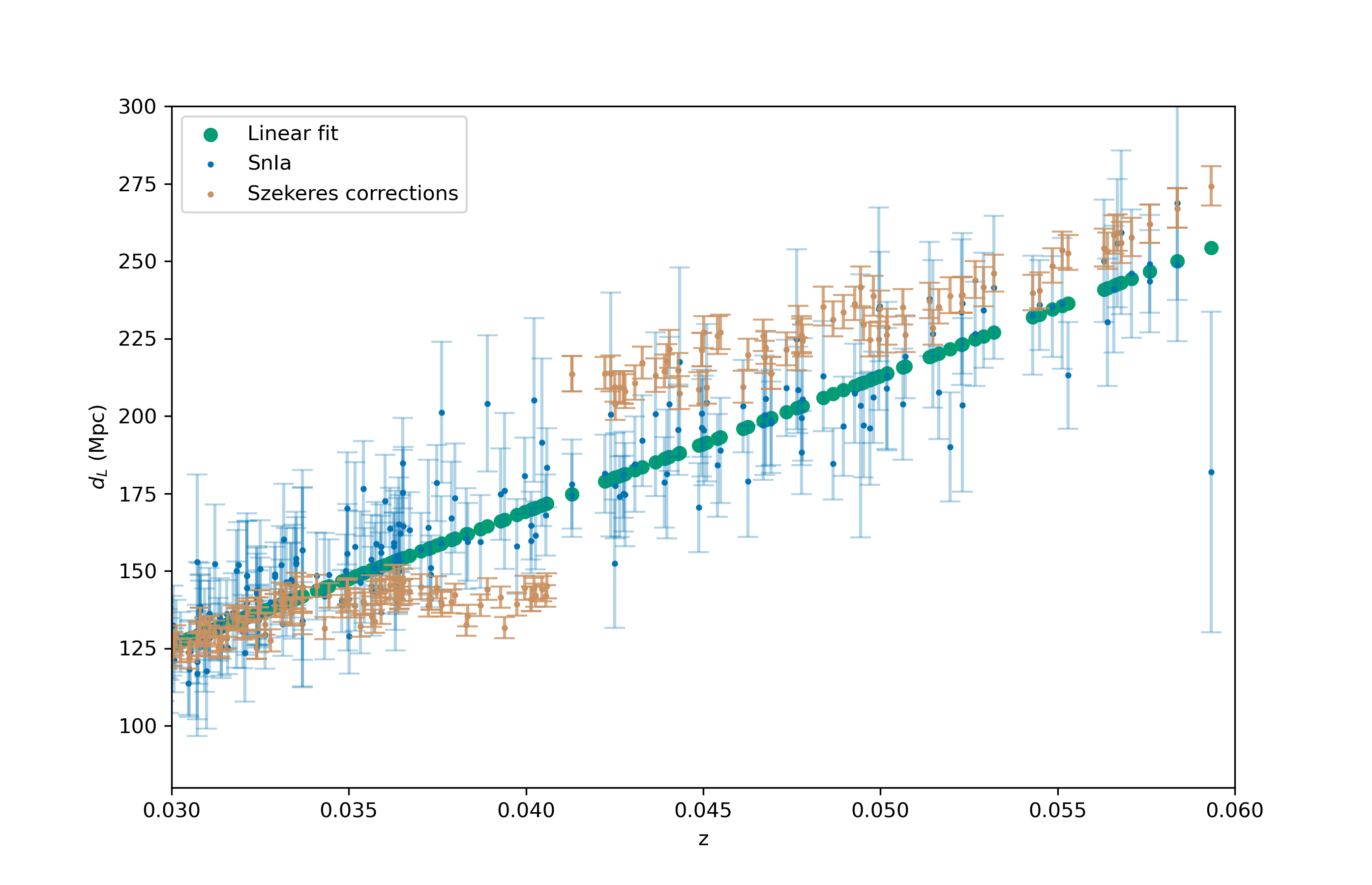}
    \caption{The Hubble diagram for the low-$z$ subsample of Pantheon+ supernovae in the range $0.023<z<0.15$ (blue), and a comparison between the linear best fit (green) and the distance corrections on individual objects predicted by our Szekeres effective model (orange).}
    \label{fig:Hubdiag}
\end{figure}

In Fig.~\ref{fig:Hubdiag} we show the comparison of the $d_L$ predictions for the SNe~Ia sample obtained from the $\Lambda$-Szekeres reconstruction (averaged over all the realisations), a fiducial FLRW cosmology\footnote{
Here, the fiducial distances in the FLRW cosmology for a given $z$ are computed using a standard cosmographic expansion up to second order in redshift. We fix the cosmographic parameters to $H_0 = 73~\mathrm{km\,s^{-1}\,Mpc^{-1}}$ and $q_0 = -0.45$.
}, and a PV-corrected FLRW background. Interestingly, the distribution of luminosity distances in the $\Lambda$-Szekeres description directly suggests an average infall within the Pantheon+ SNe~Ia sub-sample at redshift $z \approx 0.04$, as was also noted in~\cite{Sorrenti_2024} following a cosmographic analysis of the Pantheon+ sample. Furthermore, we see that the $\Lambda$-Szekeres predictions align with those of the PV-corrected fiducial cosmology at the lowest redshifts, while they increasingly deviate as $z$ grows, signalling the cumulative impact of inhomogeneities on the light propagation. This divergence reflects the fact that quasilocal variations in the expansion field can accumulate with distance—mirroring the anisotropic geometry of the local Universe—and thereby producing progressively stronger departures from the homogeneous FLRW expectation at higher redshifts along specific directions.

\begin{figure}[t]
    \centering  \includegraphics[width=1\linewidth]{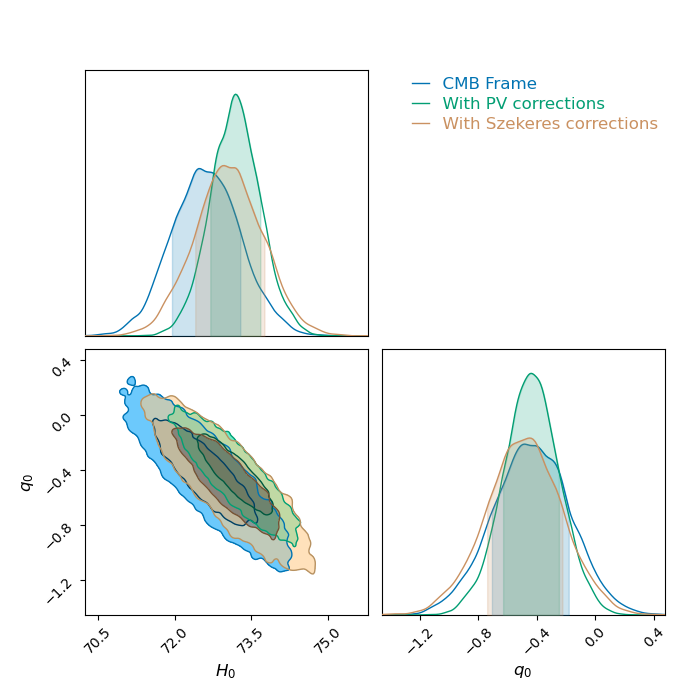}
    \caption{Posterior distribution over the cosmographic parameters $H_0$ and $q_0$ for the SH0ES supernova sample, comparing the standard FLRW model (blue contours) with the results including peculiar velocity corrections (orange contours) and the $\Lambda$-Szekeres corrections averaged over realisations (green contours).}
    \label{fig:H0q0}
\end{figure}

Then, to assess the impact of the $\Lambda$-Szekeres corrections on the local inference of $H_0$, we note that $\bar d_L(z)$ depends on the cosmographic parameters $(H_0, q_0)$ --- at the considered order in the expansion --- and thus determining how these corrections bias the local estimation of $H_0$ requires a full likelihood exploration based on the observational data. We therefore perform a Markov Chain Monte Carlo (MCMC) analysis in the $(H_0, q_0)$ parameter space using a standard $\chi^2$ log-likelihood, i.e.,
\begin{equation}
\log \mathcal{L} = -\frac{1}{2}
\left[d_L^{\mathrm{obs}} - d_L^{\mathrm{model}}\right]^{T}
\mathrm{Cov}^{-1}
\left[d_L^{\mathrm{obs}} - d_L^{\mathrm{model}}\right]\, ,
\end{equation}
where $\mathrm{Cov}^{-1}$ is the inverse covariance matrix, $d_L^\mathrm{obs} = 10^{(\mu - 25)/5}$ represents the luminosity distance obtained from the observed distance moduli $\mu$. For the  theoretical luminosity distances we consider two models, namely (i) a background, fiducial FLRW model, and (ii) the one predicted by the $\Lambda
$-Szekeres description averaged over its many realisations, i.e.,
\begin{equation}
d_L^{\mathrm{model}}(z,\hat{\mathbf{n}})=
\begin{cases}
\displaystyle
\frac{cz}{H_0}\left[1+\frac{1}{2}(1-q_0)z\right] =: \bar{d}_L^{(2)}(z)\,,\\
\displaystyle
\bar{d}_L^{(2)}(z)\bar{\Delta} d_L^{\Lambda\text{-Sk}}(z, \hat{\mathbf{n}})=:\bar{d}_L^{\Lambda-\mathrm{Sk}}(z,\hat{\mathbf{n}})\, .
\end{cases}
\end{equation}
We perform the analysis using three covariance matrices: (i) the standard SH0ES covariance without PV corrections; (ii) the SH0ES covariance including PV corrections from the 2M++ velocity–field reconstruction; and (iii) the former augmented by adding in quadrature the standard deviation of the $\Lambda$-Szekeres corrections, $\sigma_{\Delta d_L}^\mathrm{\Lambda-Sk}$. The resulting constraints\footnote{Here, we monitor the convergence following the prescription of \texttt{emcee}, estimating the autocorrelation time $\tau$ every 100 steps and declaring convergence once $\tau$ changes by less than 1\%.} are presented in Fig.~\ref{fig:H0q0}. Consistent with \cite{Riess_2022}, including PV corrections reduces the inferred distances and increases the $H_0$ best fit from $H_0 = 72.46 \substack{+0.82 \\ -0.51}$ to $H_0 = 73.19 \pm 0.49$. A qualitatively similar trend is obtained when including the $\Lambda$-Szekeres corrections, yielding $H_0 = 72.9 \substack{+0.81 \\ -0.49} $. The latter result, consistent with the findings of \cite{Giani_2024a} regarding the impact of Laniakea on cosmological inference, suggests that the Hubble tension cannot be resolved solely by modelling the expansion rate field within the local Universe. On the contrary, accounting for the local expansion field --- even through a fully anisotropic treatment of the cosmic neighbourhood --- appears to exacerbate the tension. This points toward the presence of an overall overdensity, rather than an underdensity, within the local cosmic web.

\begin{figure}
    \centering
    \includegraphics[width=1\linewidth]{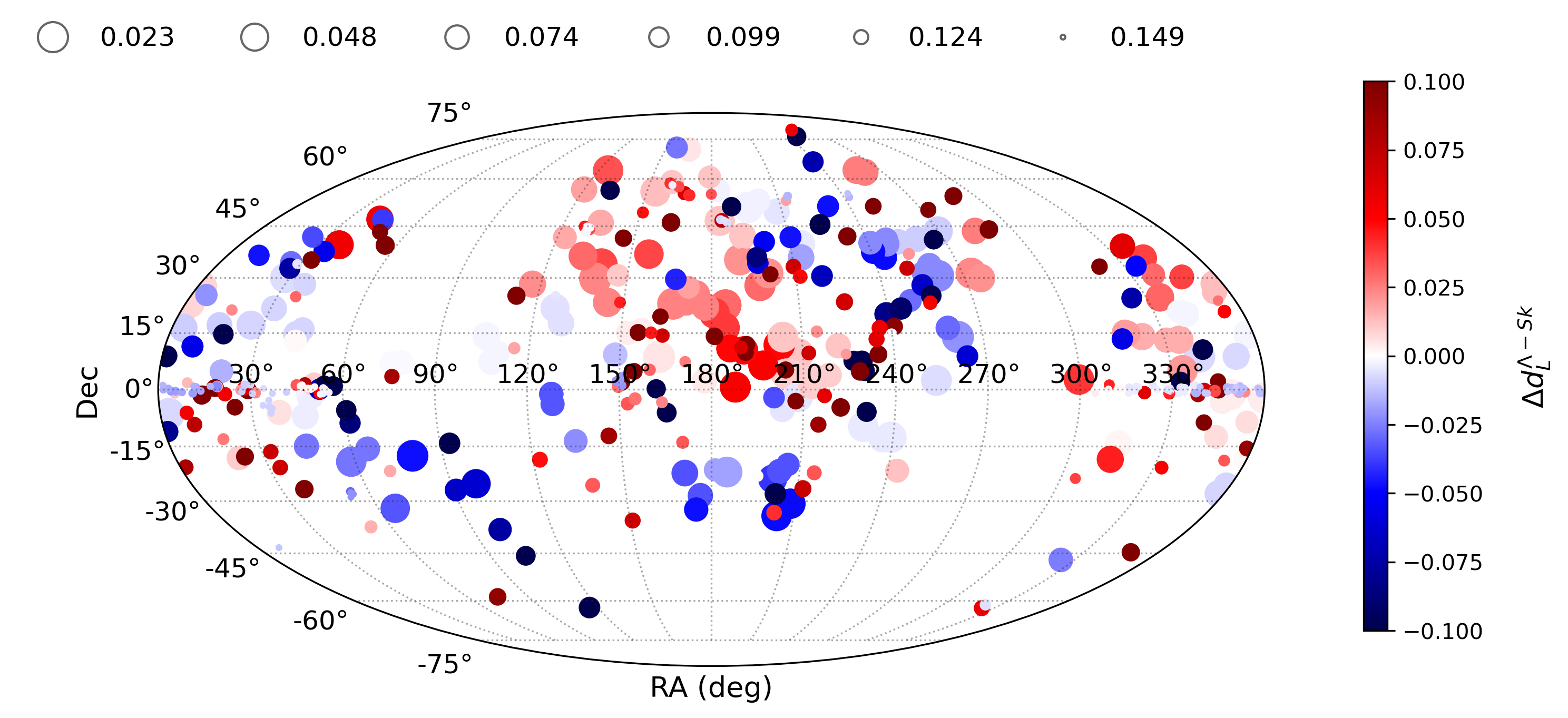}
    \caption{Sky map of the $\Lambda$-Szekeres distance corrections, $\bar{\Delta} d_L^{\Lambda\text{-Sk}}(z, \hat{\mathbf{n}})$, for the SNe~Ia within the low-redshift Pantheon+ sample, with the size of the dots indicating their redshifts in the CMB frame.}
    \label{fig:skycorr}
\end{figure}

\begin{figure}
    \centering
    \includegraphics[width=0.98\linewidth]{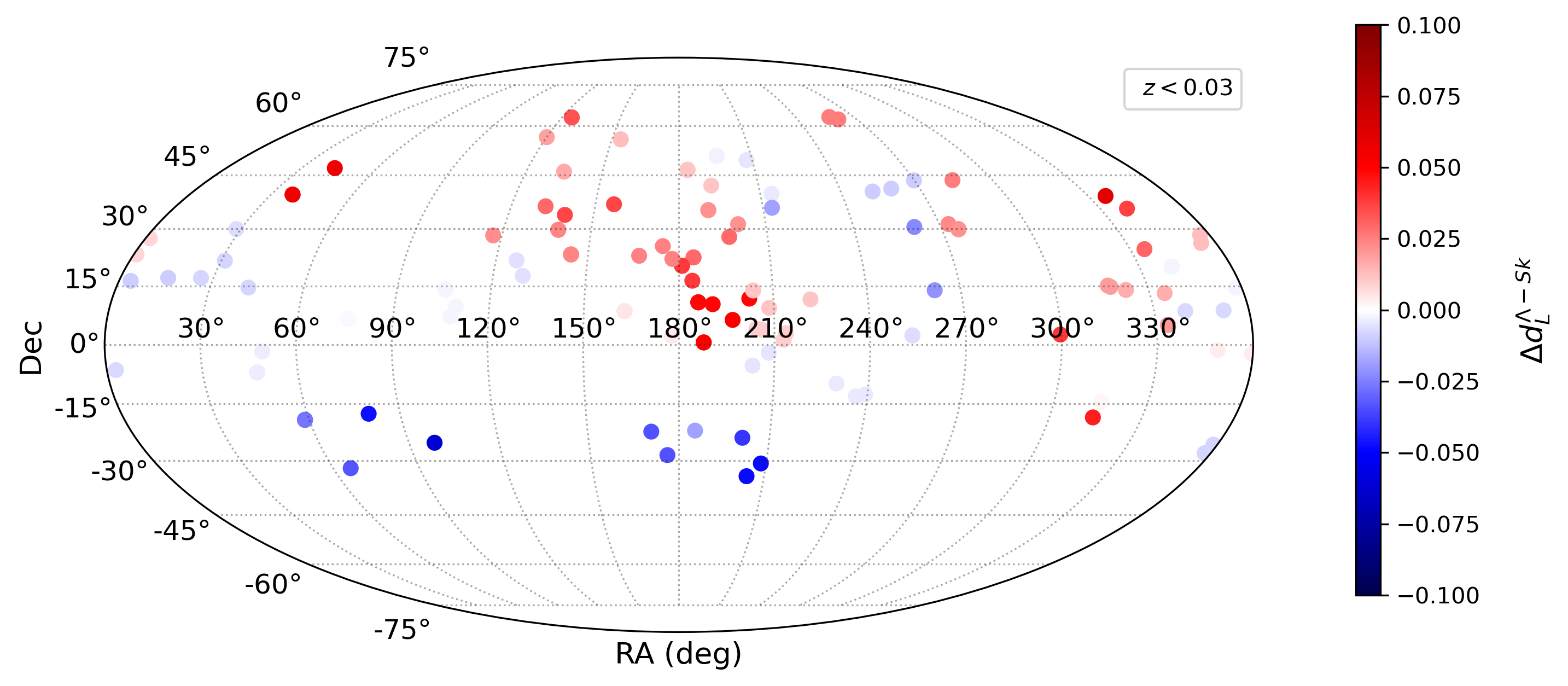}
    \includegraphics[width=0.98\linewidth]{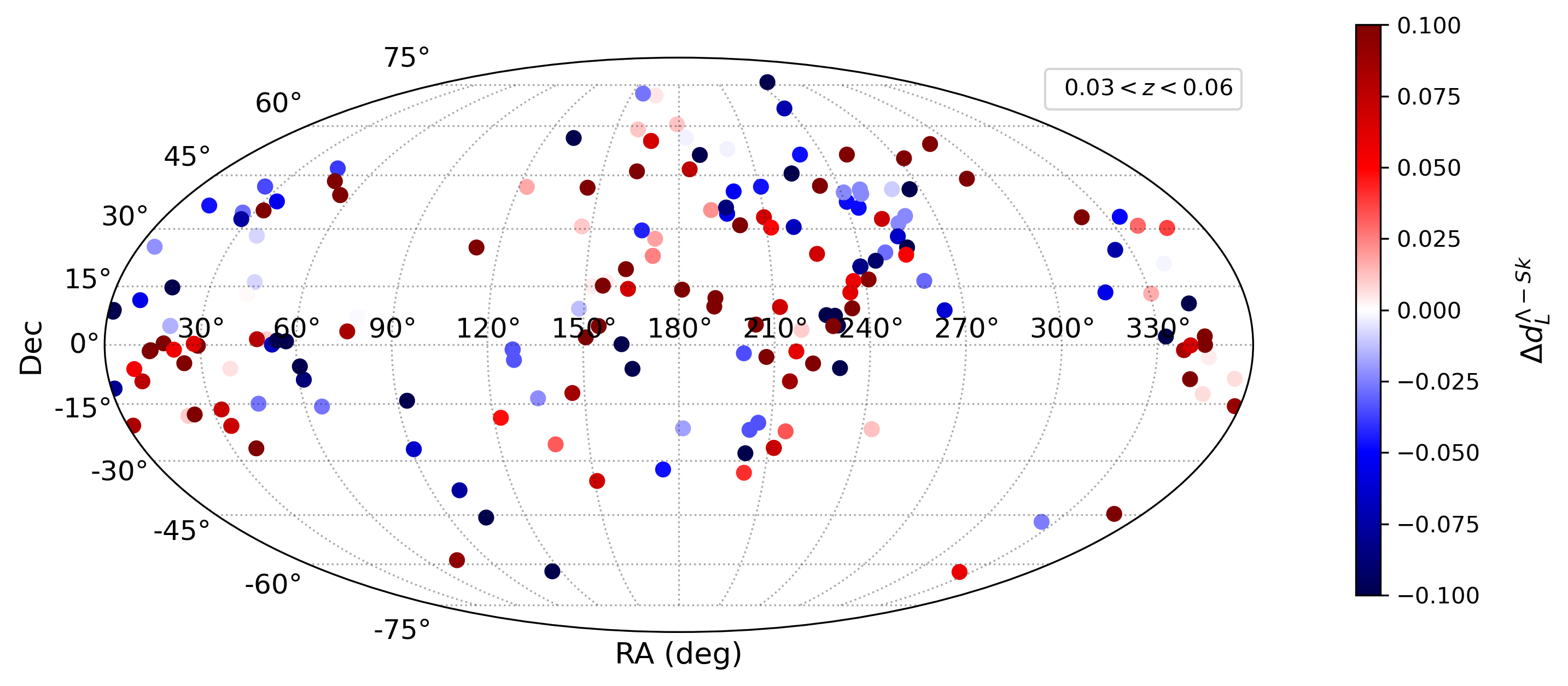}
    \includegraphics[width=0.98\linewidth]{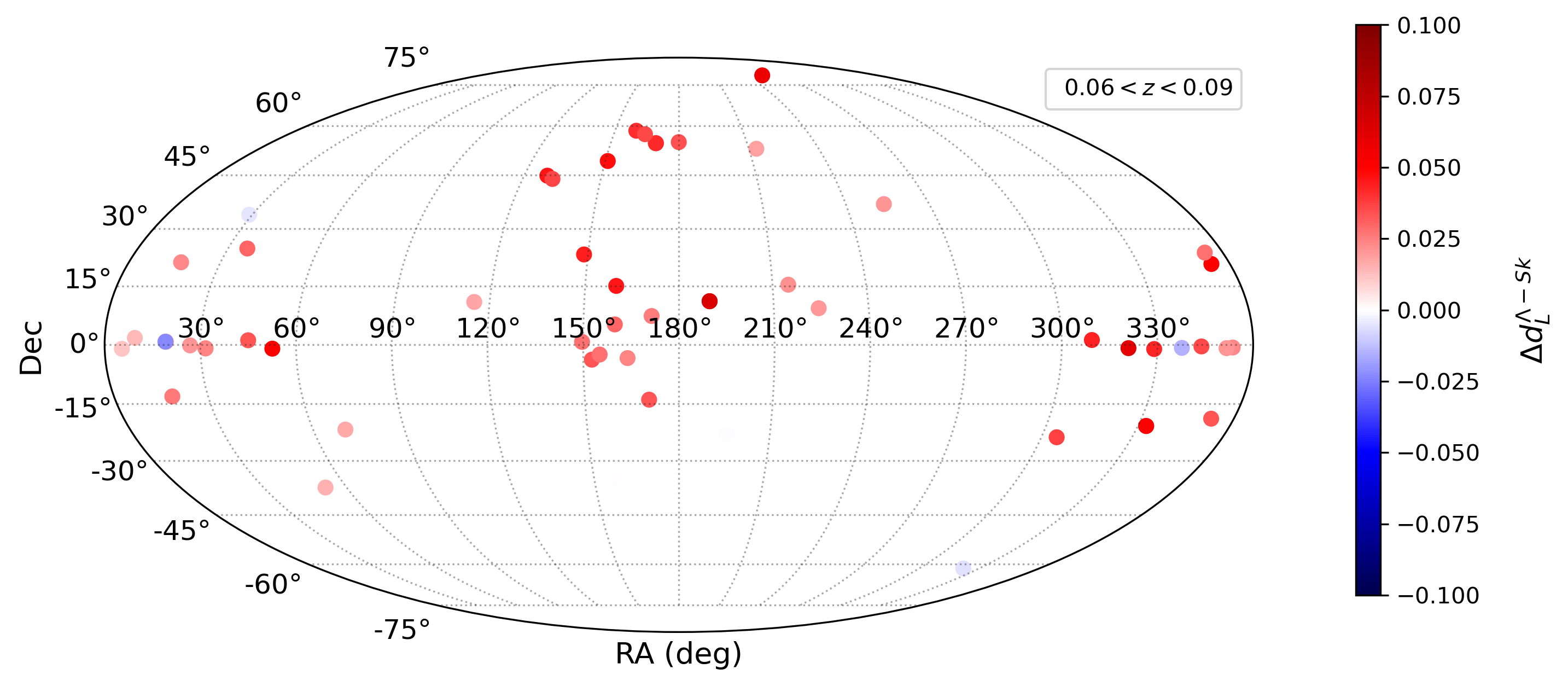}
    \caption{Sky map of the $\Lambda$-Szekeres distance corrections. $\bar{\Delta} d_L^{\Lambda\text{-Sk}}(z, \hat{\mathbf{n}})$, as in Fig.~\ref{fig:skycorr}, shown in different redshift bins. One can see that, especially at low redshifts, the corrections follow a non-random pattern across the sky.}
    \label{fig:skycorr2}
\end{figure}

\begin{figure}
    \centering
    \includegraphics[width=0.95\linewidth]{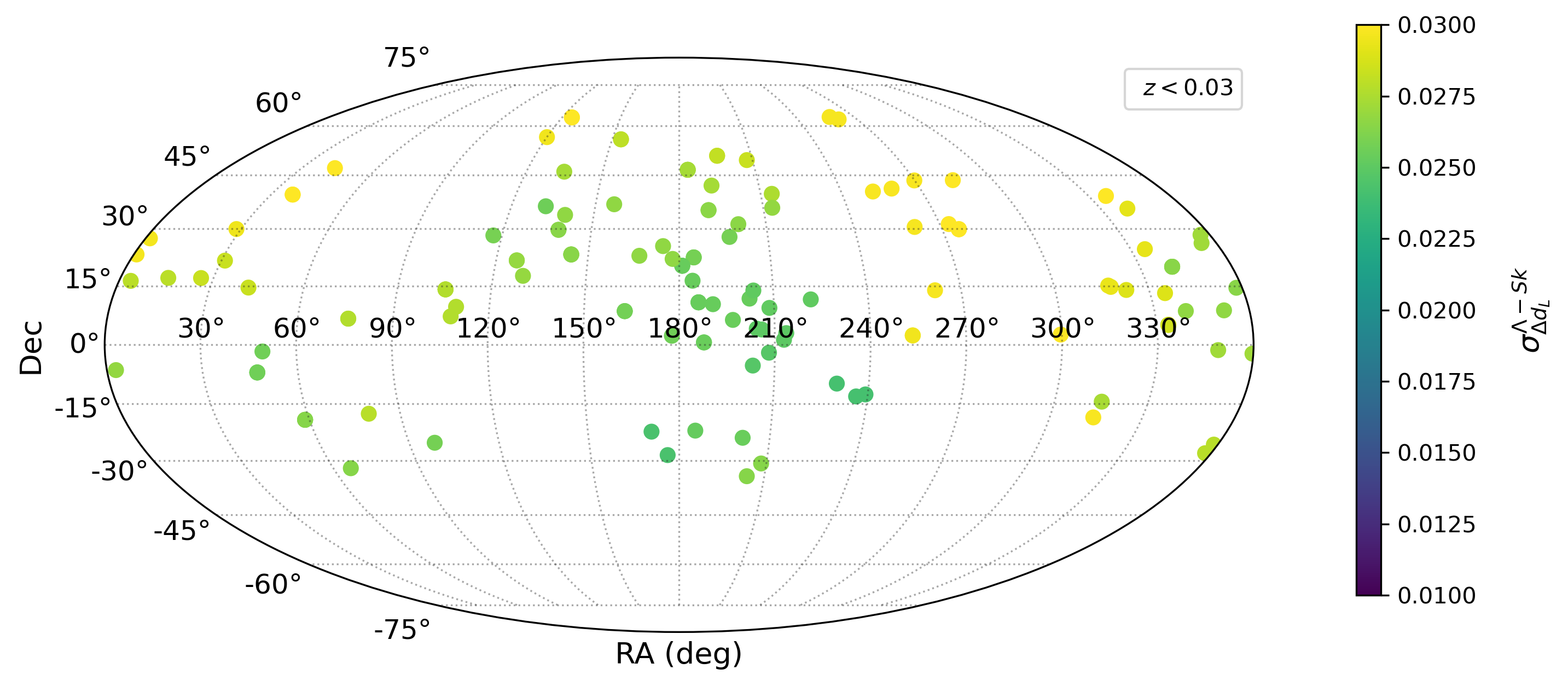}
    \includegraphics[width=0.95\linewidth]{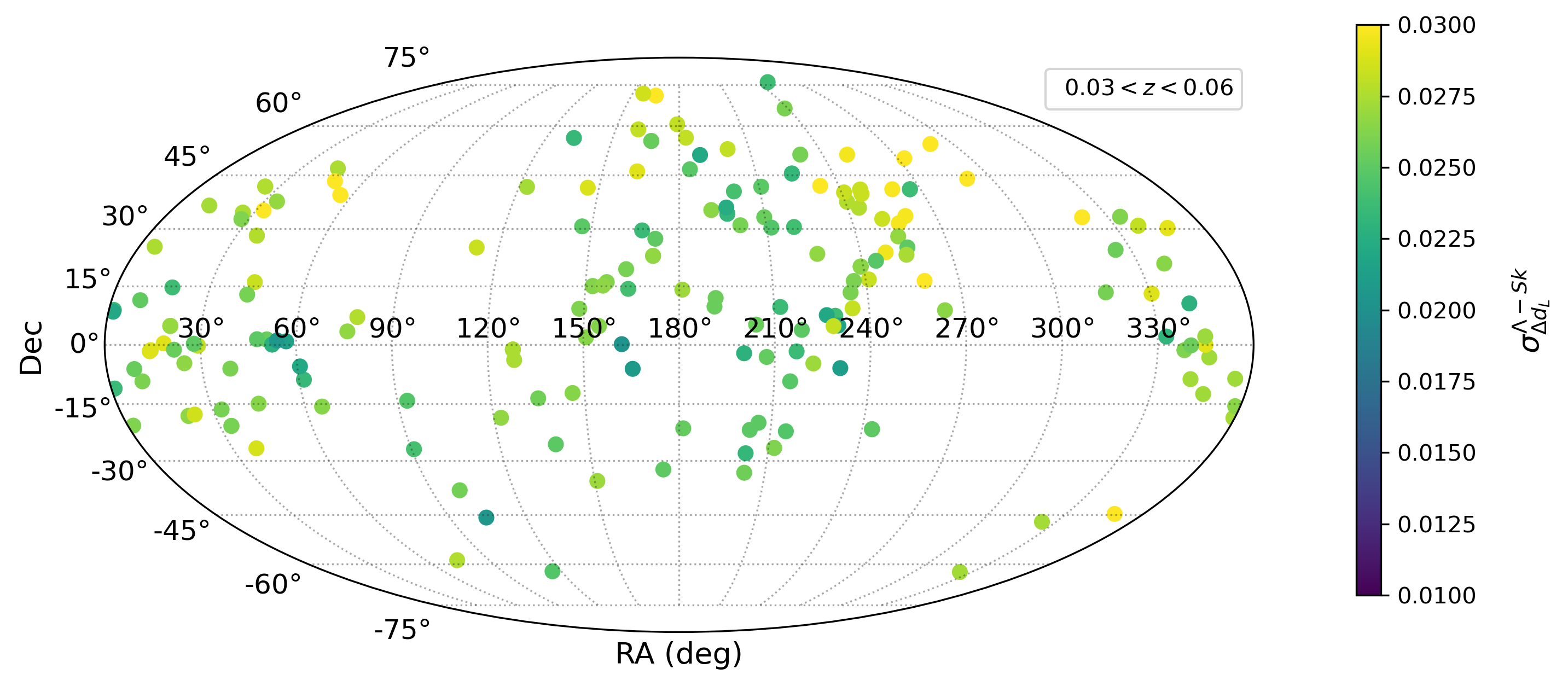}
    \includegraphics[width=0.95\linewidth]{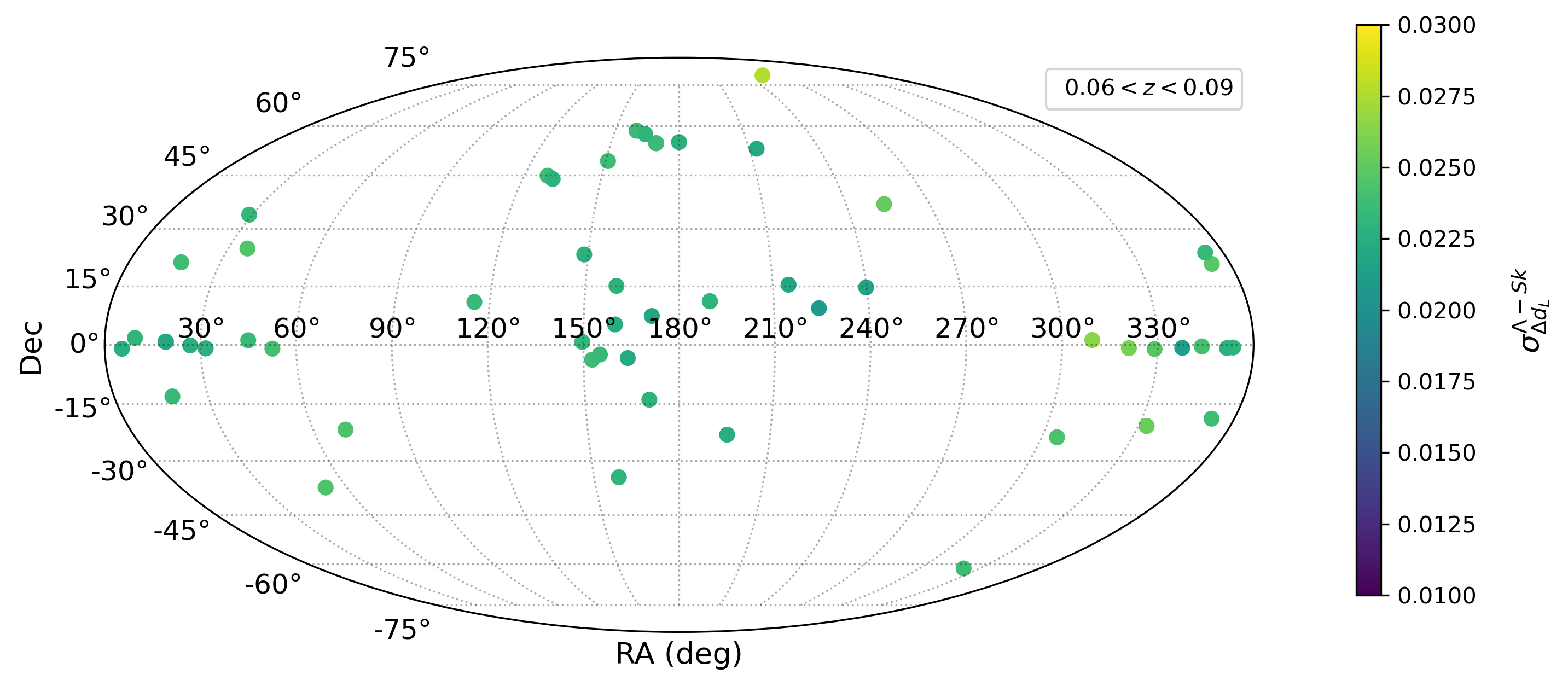}
    \caption{Sky map of the $1\sigma$ uncertainties on the $\Lambda$-Szekeres distance corrections, $\sigma_{{\Delta} d_L}^{\Lambda\text{-Sk}}(z, \hat{\mathbf{n}})$,  within the same redshift bins of Fig.~~\ref{fig:skycorr2}. No clear biases emerges with respect to the distance corrections across the redshift bins. Similarly, the sign of the predicted distance correction appears uncorrelated with the magnitude of its uncertainty.}
    \label{fig:skycorr3}
\end{figure}

Finally, it is natural to ask whether the effective–model corrections behave as random scatter or exhibit a coherent structure. As shown in Fig.~\ref{fig:skycorr}, the corrections, clearly display a directional pattern rather than a random distribution. To investigate further, in Fig.~\ref{fig:skycorr2} we show the expected distance corrections at $z = 0.03,\, 0.06,\, \text{and},\, 0.09$. These corrections, which reflect the geometry of the underlying quasilocal expansion rate field, exhibit a directional pattern that indicates an alternating arrangement of cosmic stuctures, i.e., walls and voids, within the local Universe. As such, they reveal the nontrivial geometry of the local cosmic web and its influence on distance measurements, highlighting the a priori importance of accounting for anisotropic expansion effects in cosmological analyses within the cosmic neighbourhood. This interpretation is further supported by Fig.~\ref{fig:skycorr3}, where we show the dispersion $\sigma_{\Delta d_L}^\mathrm{\Lambda-Sk}$ of the inferred distance corrections: no clear bias is found either with respect to the sign of the correction or across redshift bins. \rev{In other words, while the mean corrections trace a coherent angular and radial structure, their associated uncertainties do not exhibit either a systematic redshift-dependent trend or a clear direction-dependent bias. This indicates that the signal is not primarily driven by redshift-dependent error systematics or by the partially anisotropic sky sampling of the CF4 data, although the latter can still affect the reconstruction by driving poorly constrained or unsampled regions towards the assumed $\Lambda$CDM prior. Rather, the recovered signal reflects the geometry of the reconstructed local expansion field, and hence the local cosmic environment encoded in the CF4 data.}

\section{Conclusions}\label{sec:Conc}

Understanding how the anisotropic structure of the nearby cosmic web affects low–redshift observables is essential for precision cosmology. In this work, we addressed this issue by developing an effective $\Lambda$-Szekeres modelling of the local Universe, using a patchwork of multi-structured solutions as a proof-of-concept for a fully relativistic framework in which to describe the highly anisotropic geometry of our cosmic neighbourhood.

To constrain our model, we adopt a data–driven strategy based on the HAMLET CF4 reconstruction~\cite{Valade_2022,Valade_2024} of the local density contrast and the full three–dimensional peculiar velocity field out to redshift~$z \lesssim 0.15$~\cite{Tully_2023}. To suppress spurious small–scale nonlinearities, we smooth the original fields using a coarse–graining procedure over neighbouring voxels, ensuring that the resulting maps remain fully within the quasilinear regime. We then leverage the quasilocal formalism of the $\Lambda$–Szekeres solutions~\cite{Sussman_2012} to fit an approximate model to these smoothed data, and extract the quasilocal averages of both the expansion field and the spatial curvature in the local Universe (see Figs.~\ref{fig:Hq_all_planes} and~\ref{fig:Kq_SGXSGY}).

We found that the quasilocal average dominates the local expansion field, with deviations from it contributing only higher-order corrections. This allowed us to compute directional corrections to the luminosity distance inferred by an observer in the Milky Way, directly from the quasilocal averages within the effective $\Lambda$–Szekeres description. These corrections, driven by the inhomogeneous and anisotropic expansion encoded in the effective $\Lambda$–Szekeres model, produce coherent sky patterns that reflect the anisotropic network of inhomogeneities in the local Universe, as shown in Figs.~\ref{fig:skycorr} and~\ref{fig:skycorr2}.

To assess the impact of these corrections on cosmological inference, we used the Pantheon+ SNe~Ia subsample in the redshift range $0.023 < z < 0.15$ --- the same range employed by the SH0ES collaboration to infer the Hubble constant. We performed a bootstrapping procedure over realisations of the employed CF4 reconstruction, and hence of the extracted quasilocal average expansion field, to determine for each supernova in the sample the directional corrections to the luminosity distances relative to the fiducial FLRW cosmology, along with their associated uncertainties. These corrections --- which we find can modify supernova distances by up to 10\% --- were then incorporated into an MCMC analysis of a second-order cosmographic expansion's parameter space to quantify the bias introduced by the local Universe expansion field on the estimation of $H_0$. We found that the inferred value of $H_0$ increases when considering the corrections predicted by our approximate $\Lambda$–Szekeres model, worsening rather than relieving the Hubble tension. 

These findings align with a previous analysis~\cite{Giani_2024a}, which suggests that local structures cannot reconcile early-Universe and late-Universe determinations of $H_0$. However, while~\cite{Giani_2024a} considered only a homogeneous, anisotropic model of the Laniakea supercluster, we account for the full network of filaments, clusters, and voids in the local Universe, providing a more complete assessment of how local structure affects supernova distances and, consequently, the inferred value of $H_0$. Nonetheless, we find that including the full complexity of the local cosmic web does not seem to alleviate the Hubble tension, indicating that the solution might lie beyond local-structure effects.

\rev{However, we must emphasise that in both~\cite{Giani_2024a} and in this paper, the modelling of the local Universe relies on a reconstruction of the density and velocity fields formulated within a background-plus-perturbations description. These reconstructed fields are then reinterpreted within an effective, fully relativistic framework. A natural extension would therefore be to bypass this intermediate step and constrain the effective spacetime geometry directly from the observational data. This would amount to fitting the free functions of a chosen class of exact solutions, such as $\Lambda$--Szekeres models, directly to distance and redshift data, as proposed e.g., in~\cite{Mustapha_1997,Tokutake_2016,Celerier_2026}.}

\acknowledgments 
We are grateful to Roberto A.\ Sussman for insightful comments, and engaging discussions. We also thank Francesco Bennetti, Tamara M. Davis, Christopher Harvey-Hawes, Emma Johnson, Zachary Lane, Pierre Mourier, and Shreyas Tiruvaskar for useful discussions. MG and MH recognise supports by the Marsden Fund grant M1271 administered by the Royal Society of New Zealand, Te Ap\=arangi. LG acknowledges support from the Australian Government through the Australian Research Council Centre of Excellence for Gravitational Wave Discovery (OzGrav). 

\bibliography{main}

\appendix 

\section{The quasilocal formalism for \texorpdfstring{$\Lambda$}{Λ}-Szekeres models}
\label{app:Szekeres}
The quasilocal formalism of $\Lambda$-Szekeres models, developed Sussman and Bolejko in~\cite{Sussman_2012} and expanded upon in~\cite{Sussman_2015,Sussman_2016,Sussman_2017}, has been demonstrated to be particularly well adapted to describe multi-structured $\Lambda$-Szekeres models~\cite{Sussman_2015,Sussman_2016}.  

Within this formalism, it is convenient to write the $\Lambda$-Szekeres metric in spherical coordinates as~\cite{Sussman_2012,Sussman_2015,Sussman_2016,Sussman_2017}
\begin{align}
    & g_{rr} = a^2\left[\frac{\left(\Gamma -\boldsymbol{W}\right)^2}{1 - \mathcal{K}_{q,\text{ini}}r^2}+\frac{\sin^4(\theta)}{[1+\cos(\theta)]^2}\left(\mathcal{W}^2-2\frac{1+\cos(\theta)}{\sin^2(\theta)}Z\,\boldsymbol{W}\right)\right]\, , \\
    & g_{r\theta} = \frac{a^2 r\sin(\theta)}{1+\cos(\theta)}\left(\boldsymbol{W}-Z\right) \, , \,\, g_{r\phi} = -  \frac{a^2 r\sin^2(\theta)}{1+\cos(\theta)}\boldsymbol{W}_{,\phi} \, , \\
    & g_{\theta\theta} = a^2r^2\, ,\, \, g_{\phi\phi} = a^2r^2\sin^2(\theta) \, , \, \, g_{tt} = -1\, ,
\end{align}
where $a = a(t,r)$, $\Gamma = \Gamma(t,r) = 1+ra'/a$ and $a' = \partial_ra$. $\mathcal{K}_{q,\text{ini}} = \mathcal{K}_{q,\text{ini}}(r)$ is an a priori arbitrary function of $r$, whilst the Szekeres dipole function $\boldsymbol{W}(r,\theta,\phi)$ and its magnitude $\mathcal{W}(r)$ are given by
\begin{align}
    & W(r,\theta,\phi) =- X(r)\cos(\phi)\sin(\theta)-Y(r)\sin(\phi)\sin(\theta) -Z(r)\cos(\theta) \, , \\
    & \mathcal{W}(r) = X^2(r) +Y^2(r) +Z^2(r) \, ,
\end{align}
where $X, Y$ and $Z$ are arbitrary functions of the radial variable. We can then define for the Hubble scalar, $H$, the curvature scalar, $\mathcal{K}$, and the density scalar, $\rho$, their respective \lq\lq q-scalars'' via~\cite{Sussman_2012,Sussman_2015,Sussman_2016,Sussman_2017}
\begin{equation}
    A_q(t,r) :=\frac{\int_0^r\int_0^{\pi}\int_0^{2\pi} A(t,r',\theta,\phi)\mathcal{F}(r')\sqrt{\mathcal{J}(t,r',\theta,\phi)}\dd r'\dd \theta\dd \phi}{\int_0^r\int_0^{\pi}\int_0^{2\pi}\mathcal{F}(r')\sqrt{\mathcal{J}(t,r',\theta,\phi)}\dd r'\dd \theta\dd \phi} \, ,
\end{equation}
where $A = \rho,\,H,\,\mathcal{K}$,  $\mathcal{F}(r) = \sqrt{1-\mathcal{K}_{q,\text{ini}}(r)r^2}$, and $\mathcal{J}$ is the determinant of the spatial part of the Szekeres metric, i.e., 
\begin{equation}
    \mathcal{J}(t,r,\theta,\phi) = a^6\frac{(\Gamma - \boldsymbol{W})^2}{1 - \mathcal{K}_{q,\text{ini}}r^2}r^2\sin^2(\theta) \, .
\end{equation}
Therefore, the q-scalars are proper-volume averages of their corresponding scalars with weight factor $\mathcal{F}$, which can be interpreted as a measure of the total gravitational binding energy within the volume\footnote{This can be seen directly for spherically symmetric systems in the Newtonian limit, for which the identification is exact~\cite{Hayward_1996}.}. We can also define the \emph{exact} fluctuations of any dynamical scalar $A$ as
\begin{equation}
    D^{(A)}(t,r,\theta,\phi) := A(t,r,\theta,\phi) - A_q(t,r)\, .
\end{equation}
It can then be shown that the dynamics of the Szekeres solutions are now completely determined by the set of first-order evolution equations~\cite{Sussman_2012,Sussman_2015,Sussman_2016,Sussman_2017}
\begin{align}
    & \dot{\rho}_q = - 3\rho_qH_q \, ,\label{eq:Suss1}\\ 
    & \dot{H}_q = -H_q ^2- \frac{4\pi G}{3}\rho_q +\frac{\Lambda}{3} \,, \label{eq:Suss2}\\ 
    & \dot{\Delta}^{(\rho)} = -3\left(1+\Delta^{(\rho)}\right)D^{(H)}\,,\label{eq:Suss3}\\
    & \dot{D}^{(H)} = -2H_q  D^{(H)} - \frac{4\pi G}{3}\rho_q\Delta^{(\rho)} + 3\left(D^{(H)}\right)^2 \,,\label{eq:Suss4}
\end{align}
where we have defined $\Delta^{(\rho)} := D^{(\rho)}/\rho_q$. The evolution equations \eqref{eq:Suss1}-\eqref{eq:Suss4} are completed by the following algebraic constraints on the dynamical variables~\cite{Sussman_2012,Sussman_2015,Sussman_2016,Sussman_2017}
\begin{align}
    & H_q^2 = \frac{8\pi G}{3}\rho_q - \mathcal{K}_q +\frac{\Lambda}{3} \, ,\label{eq:Suss5}\\
    &  D^{(\mathcal{K})} =  \frac{8\pi G}{3}D^{(\rho)} - 2H_qD^{(H)} \, . \label{eq:Suss6}
\end{align}
Finally, a further constraint is given by 
\begin{equation}
    \dot{\mathcal{G}} = 3\mathcal{G}D^{(H)}\, , \label{eq:Suss7}
\end{equation}
where we have defined $\mathcal{G}:= (\Gamma -\boldsymbol{W})/(1-\boldsymbol{W})$.
Eqs.~\eqref{eq:Suss1}-\eqref{eq:Suss6} show the potential advantage of the quasilocal formalism over other $\Lambda$-Szekeres representations. Indeed, the quasilocal formalism allows for a direct split between LTB reference model and exact, dipole-like deviations. As such, it is particularly well-suited to describe multiple structures within the local Universe.

\section{Coarse-grained structures in CF4}
\label{app:structures}
\noindent
We apply a coarse--graining to the HAMLET CF4 fields~\cite{Valade_2022,Valade_2024} in order to suppress small--scale nonlinear contamination and reconstruction artefacts that are known to arise in linear Eulerian frameworks (e.g., biased amplitude in the density contrast and locally unphysical values in voids). Concretely, we replace the field value at a given location in the original reconstruction grid with a sliding--window average over a cube of $16\times16\times16$ neighboring cells, i.e.,\ over a top--hat kernel of linear size $\simeq 64~{\rm Mpc}/h$. This operation acts as an explicit low--pass filter: fluctuations on scales below $64~{\rm Mpc}/h$ are strongly attenuated, while the large--scale gradients (which dominate the quasilocal averages used in the $\Lambda$--Szekeres calibration) are retained. In practice, this brings the density contrast into the quasi--linear range required by our effective modelling, while smoothing the velocity field into a coherent streaming flow pattern consistent with the intended large--scale description.

\begin{figure}[htb]
    \centering
    \makebox[\textwidth][c]{\hspace*{-0.023\textwidth}\includegraphics[width=1.1\textwidth]{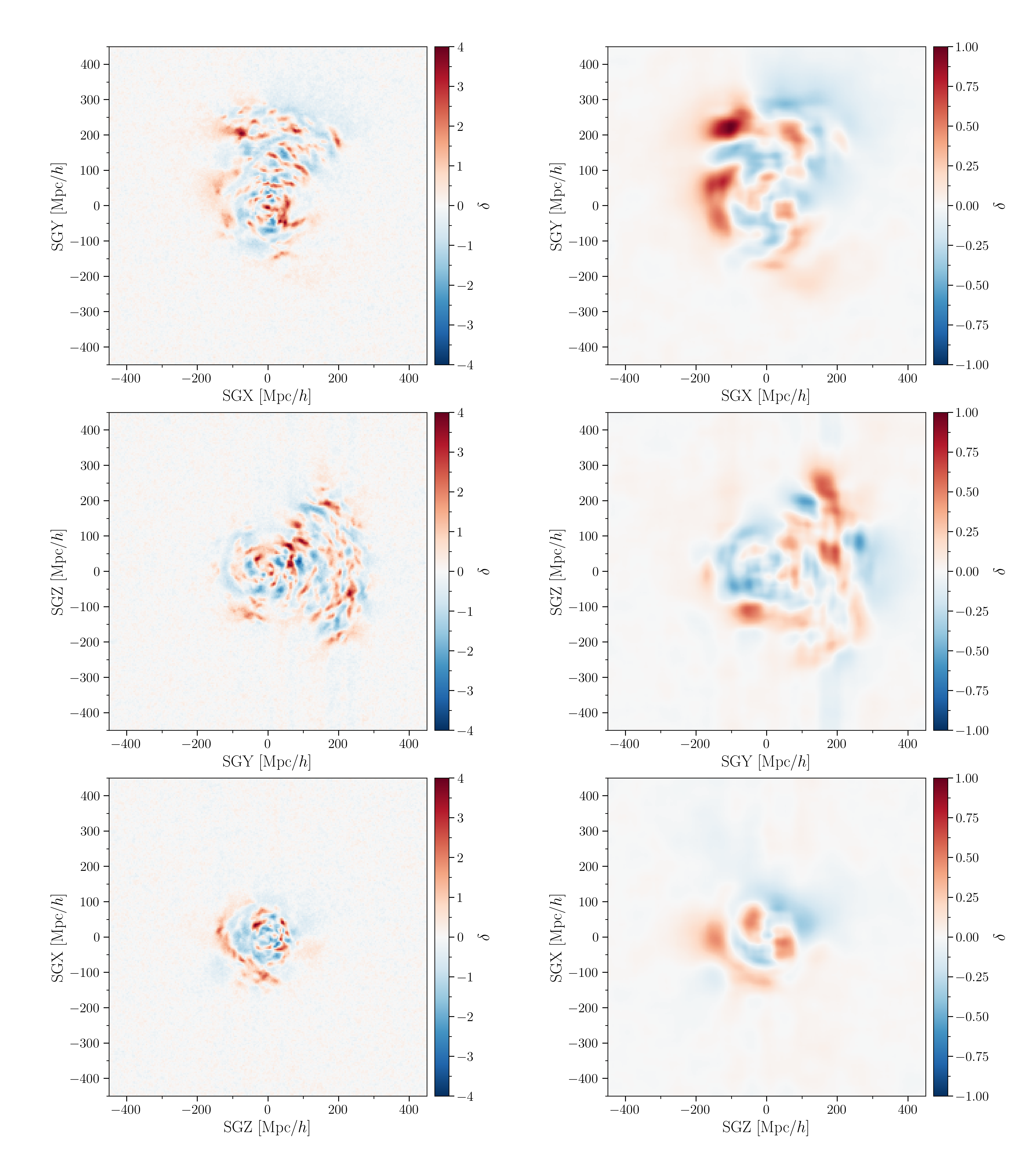}}
    \caption{Density contrast field, $\delta$, as given in the CF4 Hamlet reconstruction (left column) and the corresponding result when employing the 16-cell coarse-graining procedure (right column). The top row shows the SGX-SGY plane, the middle row the SGY-SGZ plane and the bottom row the SGZ-SGY plane.}
    \label{fig:DeltaCG}
\end{figure}

\begin{figure}[htb]
    \centering
    \makebox[\textwidth][c]{\hspace*{-0.023\textwidth}\includegraphics[width=1.1\textwidth]{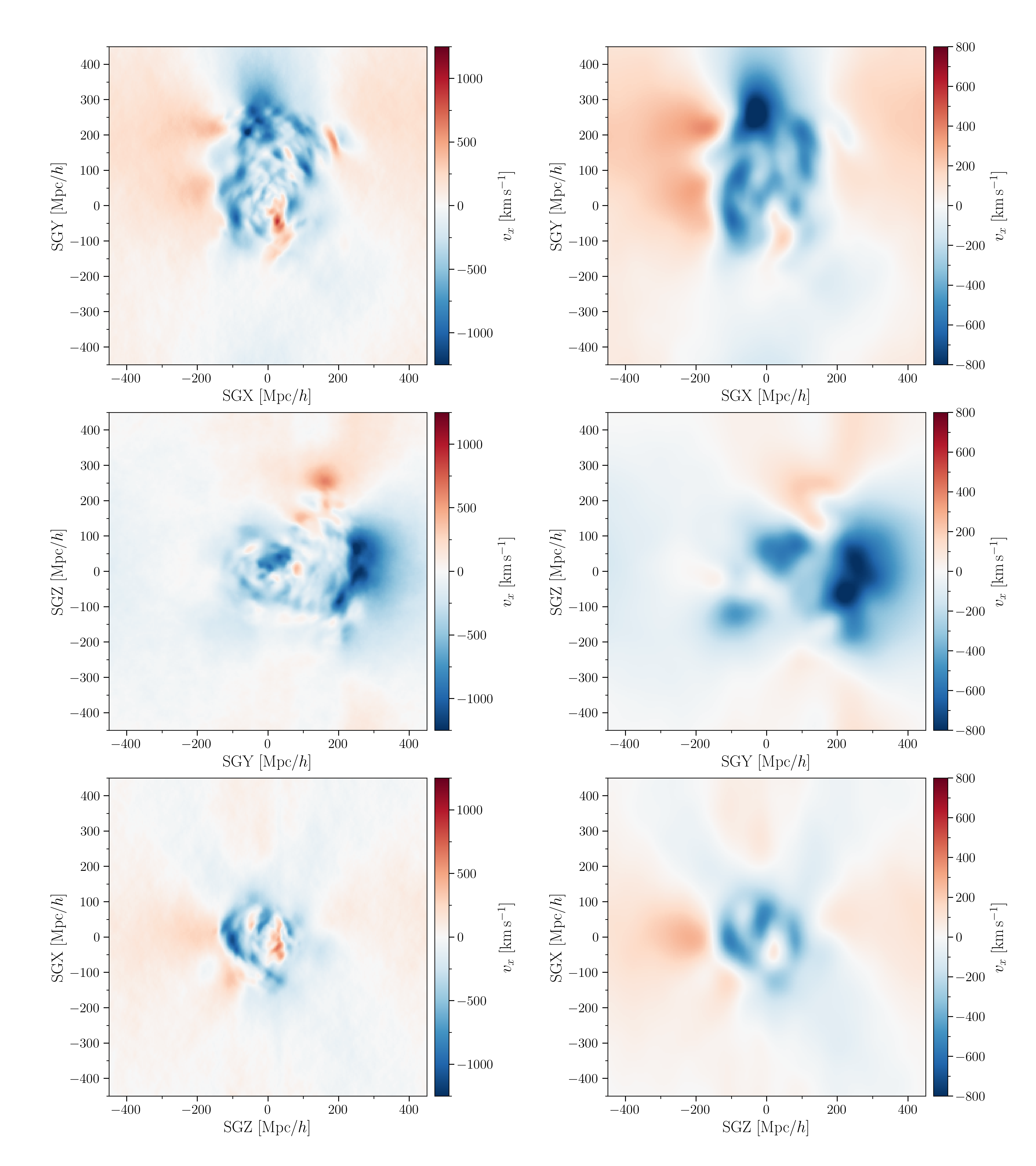}}
    \caption{SGX component of the PV field, $v_x$, as given in the CF4 Hamlet reconstruction (left column) and the corresponding result when employing the 16-cell coarse-graining procedure (right column). The top row shows the SGX-SGY plane, the middle row the SGY-SGZ plane and the bottom row the SGZ-SGY plane.}
    \label{fig:VxCG}
\end{figure}

\begin{figure}[htb]
    \centering
    \makebox[\textwidth][c]{\hspace*{-0.023\textwidth}\includegraphics[width=1.1\textwidth]{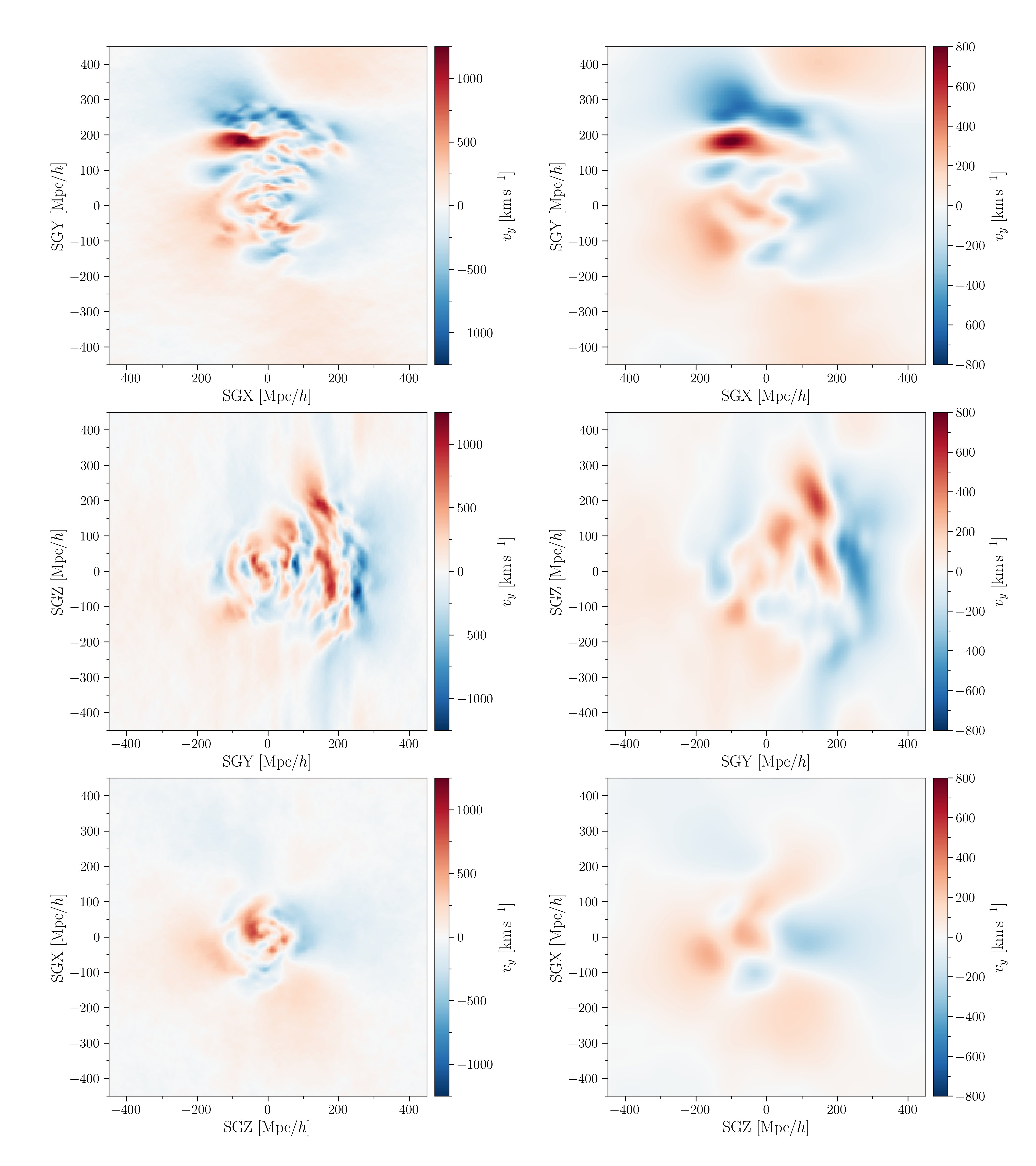}}
    \caption{SGY component of the PV field, $v_y$, as given in the CF4 Hamlet reconstruction (left column) and the corresponding result when employing the 16-cell coarse-graining procedure (right column). The top row shows the SGX-SGY plane, the middle row the SGY-SGZ plane and the bottom row the SGZ-SGY plane.}
    \label{fig:VyCG}
\end{figure}

\begin{figure}[htb]
    \centering
    \makebox[\textwidth][c]{\hspace*{-0.023\textwidth}\includegraphics[width=1.1\textwidth]{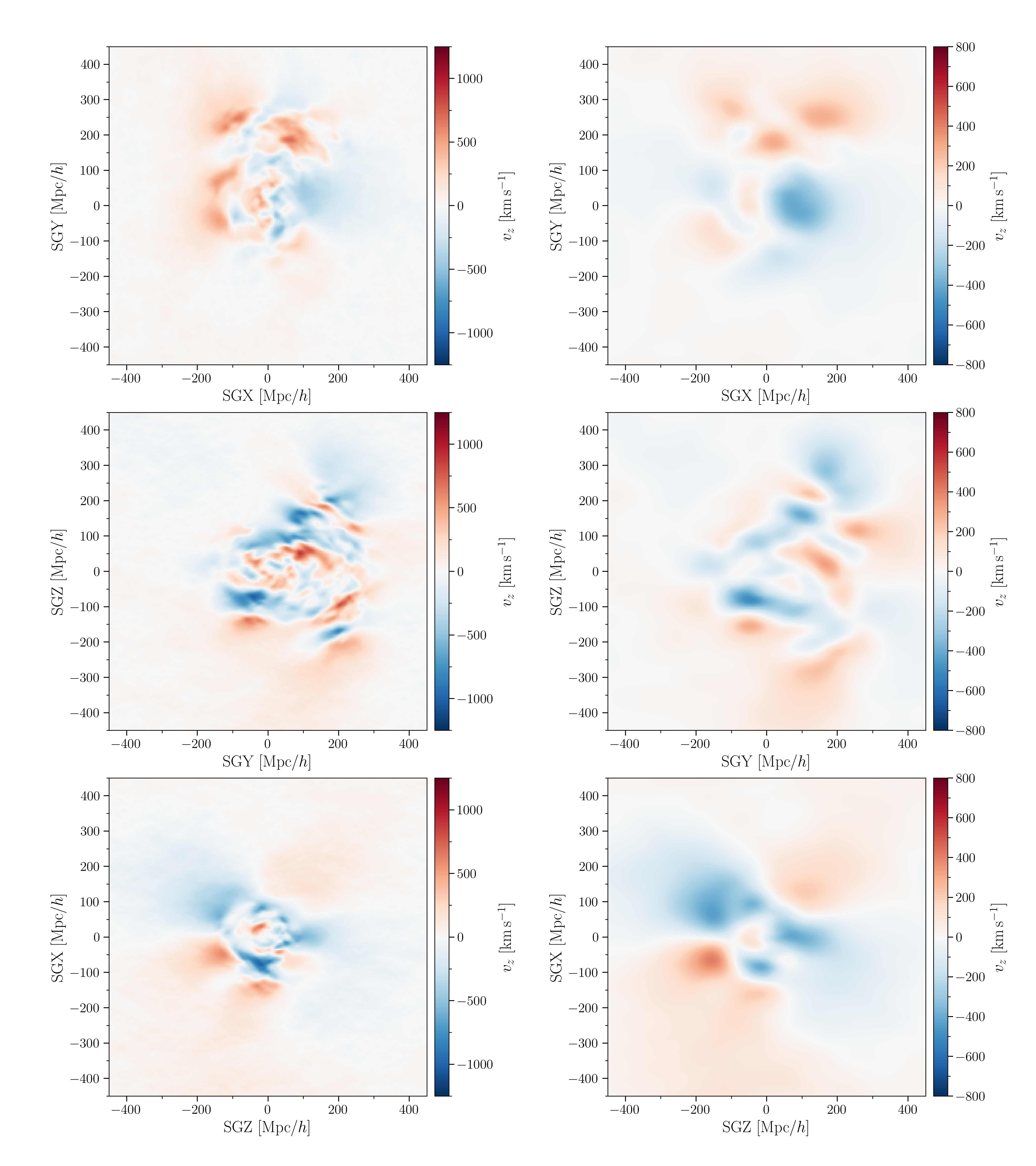}}
    \caption{SGZ component of the PV field, $v_z$, as given in the CF4 Hamlet reconstruction (left column) and the corresponding result when employing the 16-cell coarse-graining procedure (right column). The top row shows the SGX-SGY plane, the middle row the SGY-SGZ plane and the bottom row the SGZ-SGY plane.}
    \label{fig:VzCG}
\end{figure}

Fig.~\ref{fig:DeltaCG}--\ref{fig:VzCG} illustrate this directly by comparing the unsmoothed and coarse--grained maps of $\delta$, $v_x$, $v_y$, and $v_z$ (left vs right columns) across the three fundamental supergalactic planes. As expected, the coarse--graining washes out small--scale clumpiness and sharp cell--to--cell variations, most evident in the velocity components, while preserving the morphology of the dominant structures (extended overdense/underdense regions and the associated coherent streaming flows) across the full volume. In other words, the procedure reduces sensitivity to sub--voxel nonlinear motions and numerical artefacts, but does not erase the long--wavelength features that set the effective environment (and hence the quasilocal expansion/curvature fields) relevant for the distance--redshift corrections. This provides a visual validation that our smoothing is aggressive only below the target scale $\sim 64\,{\rm Mpc}/h$, and conservative with respect to the large--scale structure information that the $\Lambda$--Szekeres patchwork is designed to encode.

\section{Code routine validation}\label{app:Code}
To validate our code for the $\Lambda$-Szekeres modelling, we employ a simple simulation of a toy model, the `Hubble bubble', whose analytical behaviour is well understood. Specifically, we consider a spherical region of radius $r_{b}$ with a constant linear density contrast $\delta$, embedded in an otherwise flat FLRW background. The variation of the Hubble parameter inside the bubble is related to the density contrast and to the background expansion rate by~\cite{Marra_2013}
\begin{equation}
    \frac{\delta H}{\bar{H}} = -\frac{\delta}{3} \Theta(\delta,\Omega_m)f(\Omega_m)\;,
\end{equation}
where $\bar{H}$ is the Hubble parameter outside the density contrast bubble, $\Theta \approx 1$ in the linear regime, and $f\approx \Omega_m^{0.55}$.
To validate our modelling framework, we compare the distance–redshift predictions for light sources for (i) the effective $\Lambda$-Szekeres model, (ii) Newtonian perturbation theory, and (iii) the standard FLRW model. However, to build a Hubble diagram and compare the various models predictions, we need to specify the underlying distance–redshift relationship in the Hubble bubble model.

We begin by noting that for a generic LTB model with line-element
\begin{equation}
    \dd s^2 = -\dd t^2 + \frac{R'^2(t,r)}{1-k(r)}\dd r^2 + R^2(t,r)\mathrm{d}^2\Omega\; ,
\end{equation}
we can directly employ the redshift as a parameter along null geodesics, and obtain the radial geodesic equations~\cite{Krasinski_1997, Celerier_2024} 
\begin{align}
    \frac{dt}{dz}= -\frac{R'(r,t)}{(1+z)\dot{R}'(r,t)}\;,\; \; \;
    \frac{dr}{dz}=-\frac{\sqrt{1-k(r)}}{(1+z)\dot{R}'(r,t)}\;,
\end{align}
with the luminosity and angular distances then given by~\cite{Krasinski_1997, Celerier_2024}
\begin{equation}
    d_A = \frac{d_L}{1+z^2}=R(r,t)\;.
\end{equation}
We can then define a corresponding \lq\lq comoving" distance as the proper radial distance at a fixed cosmic time $t = t^*$ divided by the local scale factor, namely
\begin{equation}
    \chi(t^*,r):= \int_0^r \frac{R'(t^*,r)}{\sqrt{1-k(\tilde{r})}}\dd\tilde{r}\;.
\end{equation}
This can also be related to the l.o.s.\ comoving distance corresponding to an observed redshift $z$ via
\begin{equation}
    \chi(z)= \int_0^{z} \frac{c\dd z'}{(1+z')({\dot{R}'}/{R'})} \; . \label{eq:AppB_DcLTB}
\end{equation}
For the Hubble Bubble model, Eq.~\eqref{eq:AppB_DcLTB}  becomes
\begin{equation}\label{piecewiseD}
    \chi(z) = c\int_0^{\rm{min}(z,z_b)} \frac{dz'}{\bar{H}+ \delta H} + u(z-z_b)c\int_{z_b}^z \frac{dz'}{\bar{H}}
\end{equation}
where $u$ is the Heaviside step function, and $z_b$ is the redshift at which the past light cone crosses the comoving boundary of the bubble $r_b$, obtained from the relation
\begin{equation}
    r_b = c\int_0^{z_b} \frac{\dd z}{\bar{H} + \delta H}\; .
\end{equation}
Notice that in the simulation, we consider distances in the comoving grid (Mpc/h) with respect to the outer flat FLRW region, hence $h=H/100$. However, the \lq\lq true" radius should be given by $\tilde{h}=(\bar{H}+\delta H)/100$, so that
\begin{equation}
    r_b = \frac{Hr_{grid}}{H+\delta H}\;,
\end{equation}
which can be used to finally define
\begin{equation}\label{}
    z_b = (\bar{H}+\delta H)\frac{r_b}{c}\;,
\end{equation}
that together with Eq.~\eqref{piecewiseD} can be used to define our \lq\lq truth" for the Hubble diagram. To validate the code routine, we thus test the comoving distance predictions of the various effective models with respect to the \lq\lq truth" of the underlying Hubble bubble model.

\begin{figure}[htbp]
    \centering
    \begin{minipage}{0.49\textwidth}
        \centering
        \includegraphics[width=\linewidth]{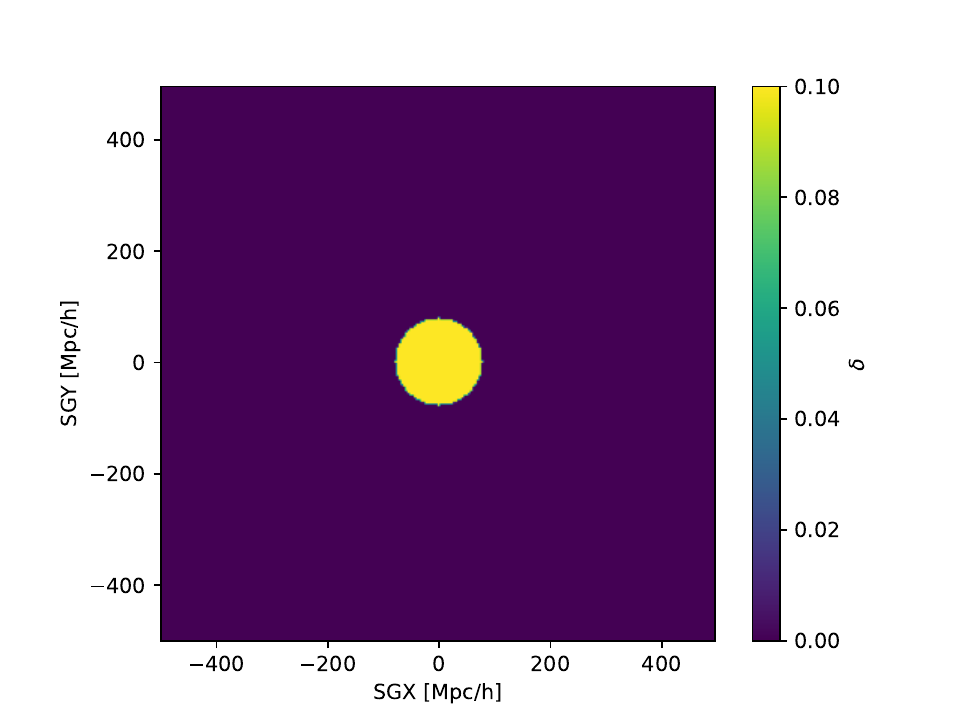}
    \end{minipage}
    \hfill
    \begin{minipage}{0.49\textwidth}
        \centering
        \includegraphics[width=\linewidth]{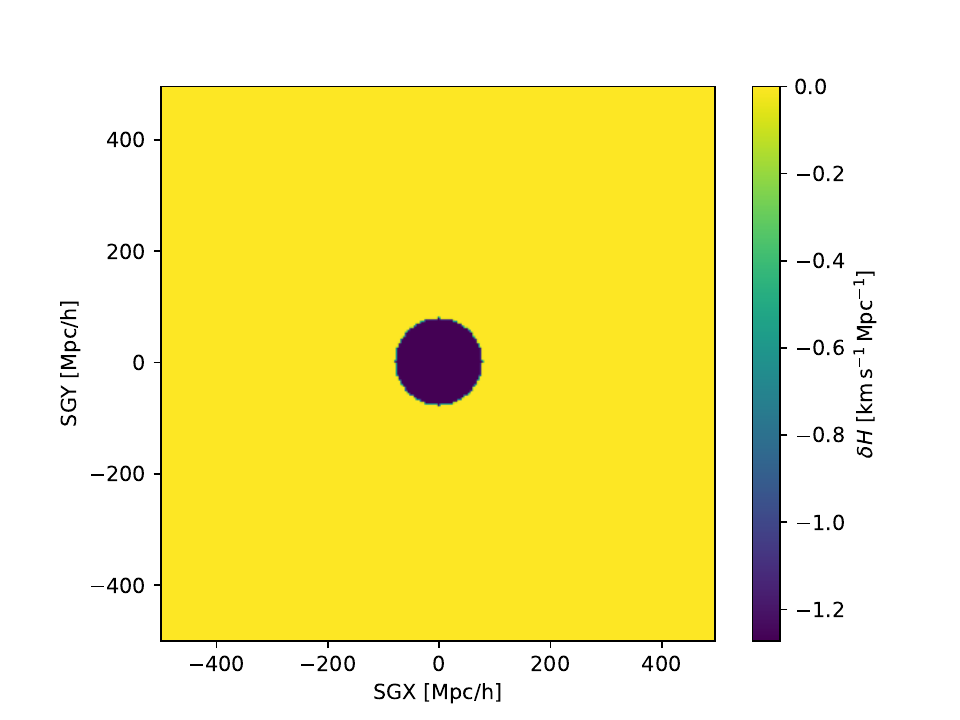}
    \end{minipage}
    \vspace{0.05cm} 
    \begin{minipage}{0.49\textwidth}
        \centering
        \includegraphics[width=\linewidth]{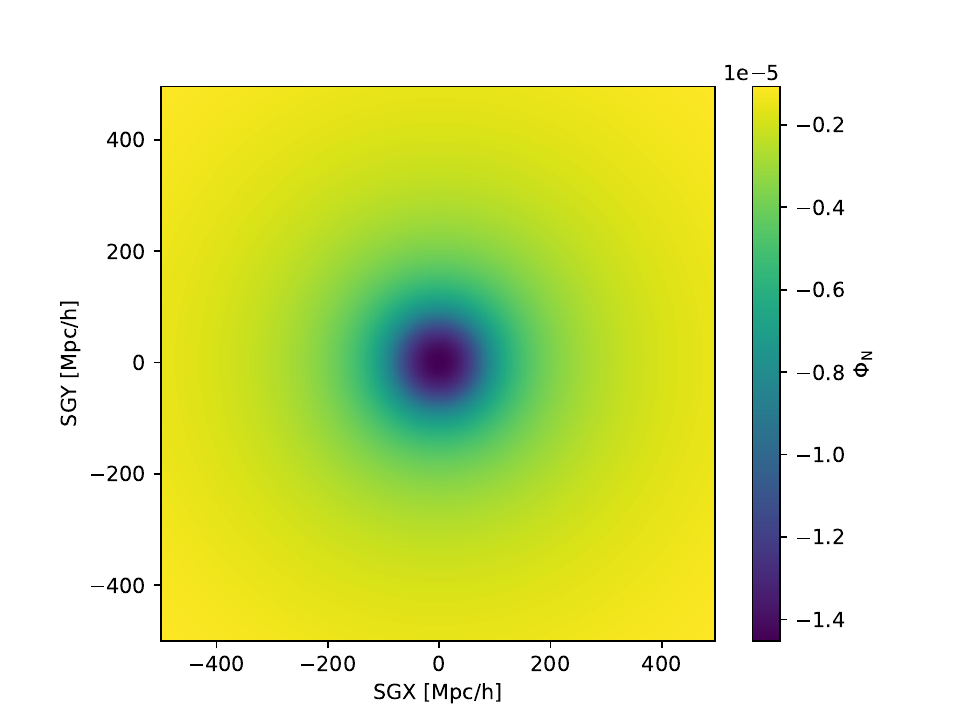}
    \end{minipage}
    \hfill
    \begin{minipage}{0.49\textwidth}
        \centering
        \includegraphics[width=\linewidth]{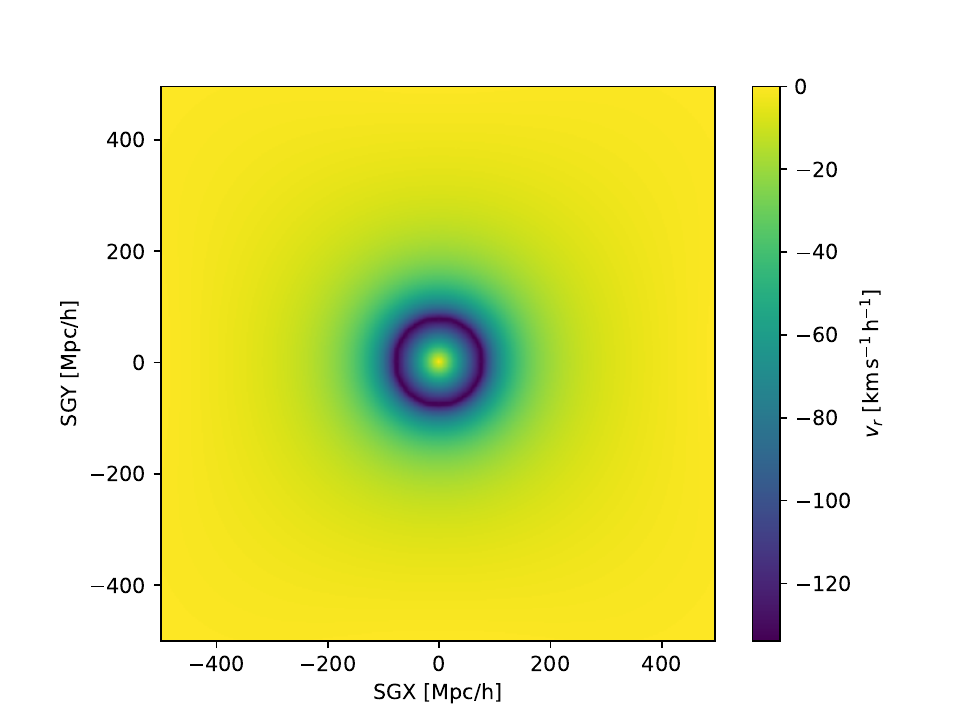}
    \end{minipage}
    \caption{Density contrast (upper-left), variation in the internal Hubble parameter (upper-right), normalised Newtonian ($\Phi_\mathrm{N}/c^2$; lower-left), and radial PVs ($v_r$, lower-right) on the SGX-SGY plane for the Hubble Bubble model simulated. The Newtonian potential and the PVs are obtained via cosmological linear perturbation theory.}
    \label{fig:appB1}
\end{figure}

Here, we consider an FLRW cosmological background with $H_0 = 74~\mathrm{km\, s^{-1}\, Mpc^{-1}}$ and $\Omega_{M0} = 0.3$, consistent with the late-time results from CF4 analyses~\cite{Duangchan_2025}. We then take a central overdensity with linear density contrast of $\delta = 0.1$, a radius within the comoving grid of $r_b \approx 80~\mathrm{Mpc/h}$, and a variation in the internal expansion rate of $\delta H \approx -1.2~\mathrm{km\,s^{-1}\,Mpc^{-1}}$. Then, from the density contrast, we directly compute the PVs and the Newtonian potential within the grid by applying linear perturbation theory, whilst imposing FLRW junction conditions at the grid boundaries (see the discussion in Subsec.~\ref{subsec:CF4}, and the reviews~\cite{KodamaSasaki_1984,Mukhanov_1992,Malik_2009}). In Fig.~\ref{fig:appB1} we show on the SGX-SGY plane the density contrast, the variation in the internal Hubble parameter, the normalised Newtonian potential ($\Phi_\mathrm{N}/c^2$), and the radial PVs within our simulation. As expected, we see that the predicted radial PVs induced by the Newtonian gravitational potential extend beyond the spherical overdensity.

We follow the procedure detailed in Sec.~\ref{sec:Fitting} to fit a $\Lambda$-Szekeres model to the simulated spherical overdensity and extract the quasilocal expansion field. We then define an approximate comoving distance for the effective $\Lambda$-Szekeres model, in accordance with the discussion in Sec.~\ref{sec:Inference}, expressed in redshift space as
\begin{equation}
    \chi_{\Lambda-\mathrm{Sk}}(z) : = c\int_0^z\frac{\dd z}{H_{q}(z)} \, .
\end{equation}
Here, we remark once again that in employing such a definition we are effectively neglecting lensing and further higher order corrections given by local deviations of the expansion field with respect to its quasilocal average. 

\begin{figure}[htbp]
    \centering
    \includegraphics[width=1\linewidth]{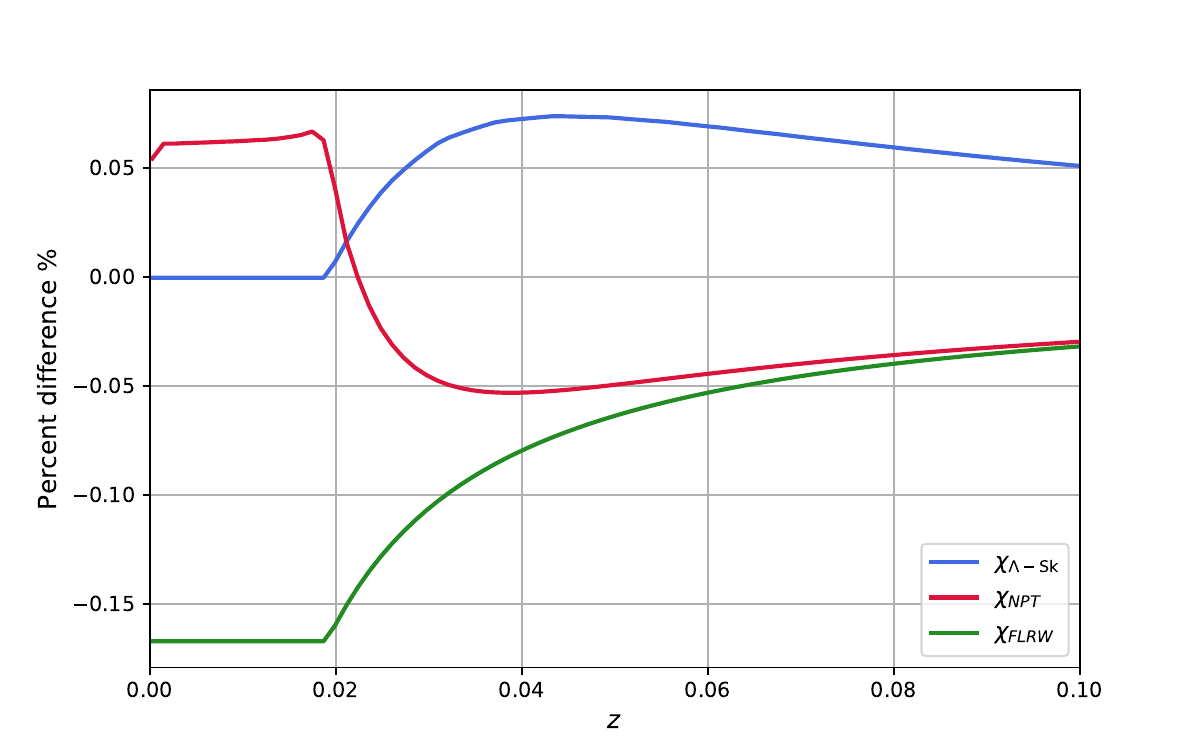}
    \caption{Percent differences, relative to the underlying Hubble-model comoving distance, of the predicted comoving distances for the $\Lambda$-Szekeres model $\chi_{\rm\scriptscriptstyle \Lambda-Sk}$, the cosmological FLRW background $\chi_{\rm\scriptscriptstyle FLRW}$, and first-order perturbation theory $\chi_{\rm\scriptscriptstyle NPT}$ around this background.}
    \label{fig:appB3}
\end{figure}

In addition to comoving distance in the $\Lambda$-Szekeres modelling, we consider the comoving distance associated to the pure FLRW background model, i.e.,
\begin{equation}
    \chi_{\mathrm{FLRW}}(z) : = \frac{cz}{\bar{H}} \, ,
\end{equation}
as well as the one predicted by considering PV and gravitational redshift corrections, in accordance to Newtonian perturbation theory, namely
\begin{equation}
    \chi_{\mathrm{NPT}}(z) : = \frac{cz}{\bar{H}} - \frac{v_r(z)} {\bar{H}} - \frac{c\Delta\Phi_\mathrm{N}(z)}{\bar{H}} \, ,
\end{equation}
where $\Delta\Phi_\mathrm{N}(z) : = \Phi_\mathrm{N}(0) - \Phi_\mathrm{N}(z)$, and the Newtonian potential is normalised with respect to $c^2$. In Fig.~\ref{fig:appB3} we show the percent difference, as a function of the redshift, of the various approximate comoving distances with respect to the \lq\lq truth" of the Hubble bubble model. 

Here, we find that the $\Lambda$-Szekeres reconstruction predicts comoving distances that are always within 0.1\% of the true values in the Hubble bubble model. Moreover, within the overdensity, the two coincide exactly, with the $\Lambda$-Szekeres reconstruction outperforming Newtonian linear perturbation theory in distance estimates, while remaining essentially comparable outside the density perturbation. Interestingly, we observe that while both the FLRW background and the linearly perturbed predictions underestimate the true distance, the $\Lambda$-Szekeres reconstruction generally slightly overestimates it.

To conclude, we have simulated the Hubble bubble —-- a fully analytical, nonlinear effective model for overdensities over an FLRW background —-- and compared the comoving distance predictions for luminous sources obtained with our $\Lambda$-Szekeres reconstruction code to those of the model itself, as well as to the pure FLRW background and linear perturbation theory around it. We find that the predictions from our code closely match the true distances, outperforming linear perturbation theory within the bubble while remaining comparable to it outside. Hence, the $\Lambda$-Szekeres reconstruction ultimately provides a fairly accurate method for modelling distance measures in inhomogeneous cosmological settings.

\end{document}